\documentclass[journal,twoside,web]{ieeecolor}
\usepackage{tmbe}
\usepackage{cite}
\usepackage{amsmath,amssymb,amsfonts}
\usepackage{algpseudocode}
\usepackage[ruled,vlined]{algorithm2e}
\usepackage{graphicx}
\usepackage{textcomp}
\usepackage{float}
\usepackage{adjustbox}
\usepackage{mathtools}
\usepackage{booktabs}
\SetKwInput{KwInput}{Input}                
\SetKwInput{KwOutput}{Output}              
\usepackage{stmaryrd}
\usepackage[font=footnotesize,justification=justified,belowskip=2pt,aboveskip=2pt]{caption}
\usepackage[math]{cellspace}
\usepackage[utf8]{inputenc}
\usepackage{url}
\usepackage{multirow}
\usepackage{xcolor}
\usepackage{comment}
\usepackage{hyperref}
\hypersetup{colorlinks=true,citecolor=teal,linkcolor=blue,urlcolor=blue}

\definecolor{brightcerulean}{rgb}{0.11, 0.62, 0.74}
\newcommand*{\revhl}{\textcolor{black}}

\newcommand{\Tone}{$\mathrm{T_1}$~}
\newcommand{\Ttwo}{$\mathrm{T_2}$~}

\newcommand{\Ttwon}{$\mathrm{T_2}$}
\newcommand{\Ttwosn}{$\mathrm{T_2^*}$}
\newcommand\blfootnote[1]{%
  \begingroup
  \renewcommand\thefootnote{}\footnote{#1}%
  \addtocounter{footnote}{-1}%
  \endgroup
}

\begin{document}

\title{A Tutorial on MRI Reconstruction: \\From Modern Methods to Clinical Implications}
\author{Tolga \c{C}ukur$^*$, Salman UH. Dar, Valiyeh A. Nezhad, Yohan Jun, Tae Hyung Kim, Shohei Fujita, Berkin Bilgic$^*$ \vspace{-0.4cm}
\thanks{This study was supported in part by TUBA GEBIP 2015 and BAGEP 2017 fellowships, and a TUBITAK 1001 Grant 121E488 awarded to T. \c{C}ukur, and in part by R01 EB028797, R03 EB031175, R01 EB032378, U01 EB026996, UG3 EB034875, and P41 EB030006 grants awarded to B. Bilgic. ($^*$Co-corresponding authors: Tolga \c{C}ukur, cukur@ee.bilkent.edu.tr and Berkin Bilgic, berkin@nmr.mgh.harvard.edu).}
\thanks{T. \c{C}ukur and V.A. Nezhad are with the Dept. of Electrical-Electronics Engineering and National Magnetic Resonance Research Center (UMRAM), Bilkent University, Ankara, Turkey, 06800. SUH. Dar is with the Institute for Artificial Intelligence in Cardiovascular Medicine, Dept. of Cardiology, Angiology, Pneumology, Heidelberg University Hospital and Heidelberg Faculty of Medicine, Heidelberg University, Germany. TH. Kim is with Dept. of Computer Engineering, Hongik University, Seoul, Korea. Y. Jun, S. Fujita, and B. Bilgic are with the Athinoula A. Martinos Center for Biomedical Imaging and Dept. of Radiology, Harvard Medical School, Massachusetts, USA.}
}
\maketitle
\blfootnote{\textcopyright~2025 IEEE. This is the author’s accepted version.
The final published version is available at \url{https://doi.org/10.1109/TBME.2025.3617575}.}
\begin{abstract}
MRI is an indispensable clinical tool, offering a rich variety of tissue contrasts to support broad diagnostic and research applications. Protocols can incorporate multiple structural, functional, diffusion, spectroscopic, or relaxometry sequences to provide complementary information for differential diagnosis, and to capture multidimensional insights into tissue structure and composition. However, these capabilities come at the cost of prolonged scan times, which reduce patient throughput, increase susceptibility to motion artifacts, and may require trade-offs in image quality or diagnostic scope. Over the last two decades, advances in image reconstruction algorithms—alongside improvements in hardware and pulse sequence design—have made it possible to accelerate acquisitions while preserving diagnostic quality. Central to this progress is the ability to incorporate prior information to regularize the solutions to the reconstruction problem. In this tutorial, we overview the basics of MRI reconstruction and highlight state-of-the-art approaches, beginning with classical methods that rely on explicit hand-crafted priors, and then turning to deep learning methods that leverage a combination of learned and crafted priors to further push the performance envelope. We also explore the translational aspects and eventual clinical implications of these methods. We conclude by discussing future directions to address remaining challenges in MRI reconstruction. The tutorial is accompanied by a Python toolbox ({\small \url{https://github.com/tutorial-MRI-recon/tutorial}}) to demonstrate select methods discussed in the article.
\end{abstract}
\vspace{-0.2cm}

\bstctlcite{IEEEexample:BSTcontrol}

\section{Introduction}
Magnetic Resonance Imaging (MRI) offers unparalleled diversity in contrast mechanisms to examine tissue anatomy, composition, and function \cite{Bernstein2004HandbookSequences}. Clinical MRI protocols routinely include sequences that generate anatomical images with e.g. T\textsubscript{1}, T\textsubscript{2}, T\textsubscript{2}\textsuperscript{*} and FLAIR weighting \cite{de1992mr,hajnal1992use}. These are often complemented by diffusion-weighted imaging (DWI), which plays a central role in stroke assessment \cite{moseley1990early} and cancer imaging \cite{koh2007diffusion}. Given access to higher-end hardware, research protocols further expand this repertoire using advanced sequences sensitive to magnetic susceptibility \cite{de2010quantitative,haacke2004susceptibility}, blood oxygen-level-dependent (BOLD) responses \cite{belliveau1991functional,ogawa1990oxygenation}, high-b-value diffusion \cite{setsompop2013pushing}, perfusion \cite{detre1990mrm}, or metabolite concentrations \cite{lin2007mrm}. Yet, despite this versatility, MRI remains limited by characteristically long acquisition times, often forcing undesirable trade-offs between spatial, temporal, or angular resolution, signal-to-noise ratio (SNR), anatomical coverage, and inclusion of specific contrast mechanisms \cite{fessler_review,hammernik2022physicsdriven}.

Over the past two decades, significant progress has been made toward mitigating these limitations by combining developments in scanner hardware and pulse sequences with sophisticated image reconstruction algorithms. Parallel imaging (PI) \cite{Sodickson1997,SENSE,griswold2002generalized}, leveraging undersampling of k-space acquisitions and high-channel count coil arrays \cite{phasedarray}, marked a pivotal advance by enabling accelerated scans across a broad range of applications. Yet, because MRI reconstruction from undersampled acquisitions is fundamentally an ill-posed inverse problem, recovering high-fidelity images often requires integration of external knowledge in the form of priors \cite{ravi_review,uecker_review,liang_review}. Building upon the foundation laid by PI, a first generation of methods introduced \textit{hand-crafted priors} designed by experts to constrain the solution space for plausible reconstructions \cite{fessler_review,uecker_review}. 
Prominent examples include compressed sensing (CS) methods enforcing sparsity of MR images in linear transform domains \cite{lustig2007sparse,block2007undersampled,akccakaya2010compressed,murphy2012fast,liang2009accelerating,feng2017compressed} and low-rank methods promoting structured representations in spatiotemporal or spectral dimensions of MRI data \cite{shin2014calibrationless,mani2017multi,LORAKS,ALOHA}. Although powerful, these approaches are inherently limited by the simplifying assumptions underlying their priors, which may fall short of fully capturing the rather non-trivial distribution of MR images. 

More recently, a second generation of methods has emerged that leverages \textit{learned priors}—implicit regularizers that are directly inferred from exemplar data \cite{sandino2020SPM,hammernik2022physicsdriven,chen2022review}. These priors are typically implemented as mappings through neural network architectures, which perform hierarchical (i.e., multi-layered) nonlinear transformations on the data to extract salient features and capture statistical dependencies present in MR images \cite{heckel2024deep}. Whether used standalone or in hybrid formulations with classical methods \cite{Wang2016,ADMM-CSNET,hammernik2018learning,Mardani2019b,schlemper2017deep,MoDl,quan2018compressed,Variational_end2end,Biswas2019,Dar2017,zhu2018image,nath2020datadriven}, deep learning (DL) approaches based on neural networks have driven substantial improvements in image quality and scan acceleration. However, they also introduce potential challenges—including reliance on large-scale, high-quality training data, suboptimal generalizability across scanners or anatomies, and implicit nature of learned priors that can complicate clinical validation—all of which remain active topics of ongoing research.

\begin{figure}
\centering
\vspace{-0.1cm}
\includegraphics[width=\columnwidth]{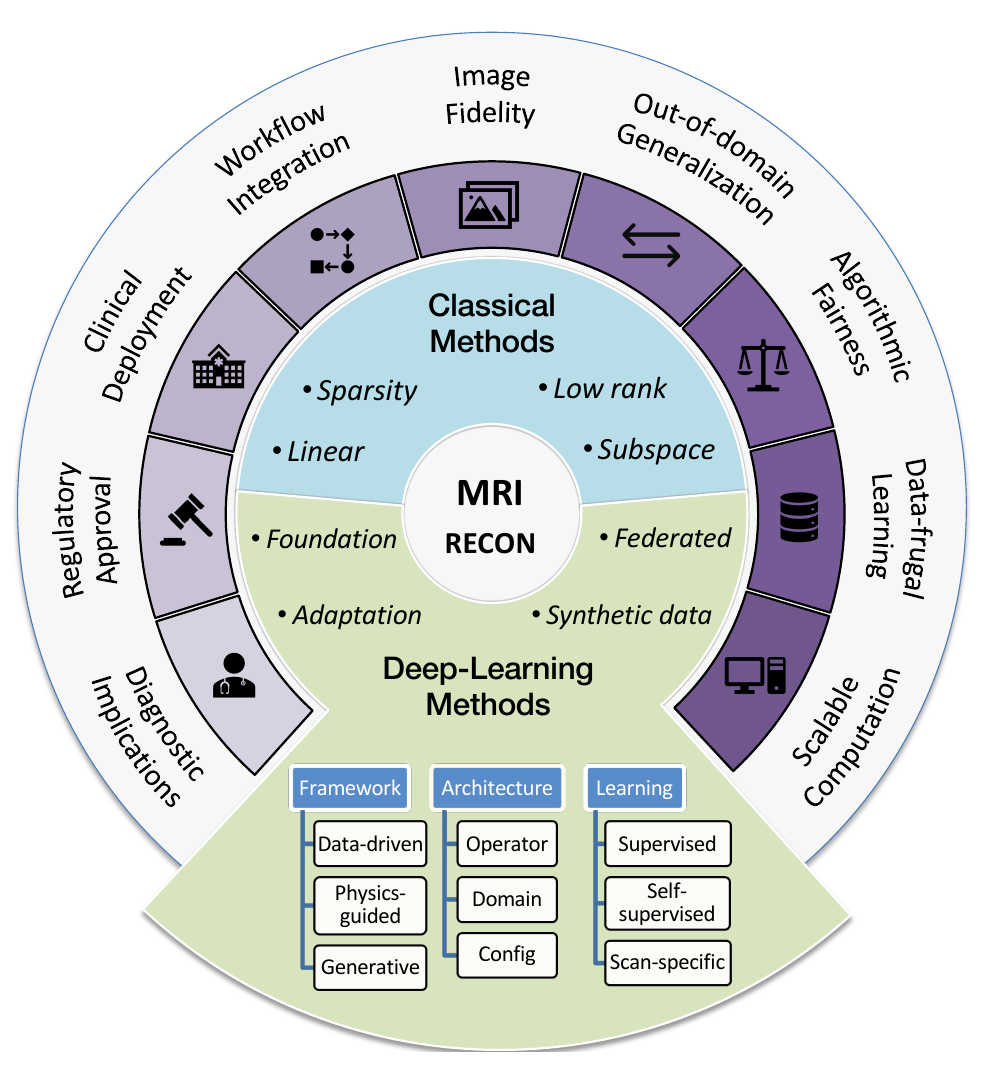}
\caption{A landscape of MRI reconstruction methods from classical approaches to modern deep learning are surveyed, alongside discussions on clinical impact and translation, as well as open challenges.}
\label{fig:grandscheme}
\end{figure}

Given the rapid pace of innovation, staying abreast of the evolving landscape of reconstruction methodology can be challenging. This tutorial is intended as an accessible introduction to MRI reconstruction—bridging foundational concepts, technical advances, and clinical implications (Fig. \ref{fig:grandscheme}). To provide a conceptual road map, we begin by revisiting core principles of spatial encoding, k-space acquisition, and the inverse problem that underlies reconstruction. We then survey both classical methods based on hand-crafted priors, and DL approaches based on learned priors. We pose a tripartite categorization to group DL methods by \textbf{(i)} the overarching framework (e.g., data-driven vs. physics-guided vs. generative), \textbf{(ii)} architectural elements (e.g., convolutional vs. attentional operator, image vs. k-space domain, single- vs. multi-contrast configuration), and \textbf{(iii)} learning paradigms (e.g., population-level vs. scan-specific). We give special attention to the clinical implications and translation of these reconstruction tools, including their potential to reduce scan times, improve image quality, and expand access to advanced imaging protocols. We conclude the tutorial with a discussion of key open challenges and future opportunities for continued innovation in the field.

\section{Foundations of MR Image Reconstruction}

\subsection{Generation of Tissue Contrast}
MRI relies on magnetization of spins in nuclei, most commonly hydrogen, under a large static field $B_0$. During imaging, the net magnetic moment in the longitudinal direction of $B_0$ is tipped to the transverse plane via a radiofrequency (RF) pulse. Magnetization then recovers to its equilibrium state with tissue-specific relaxation time constants of \Tone in the longitudinal axis, and \Ttwo in the transverse plane. Hence, the transverse magnetization, $m(\mathbf{r})$ with $\mathbf{r}=(x,y,z)$ denoting spatial location, depends on the proton density $\rho$, relaxation time constants, and the amplitude and timing of RF excitations and magnetic field gradients in the pulse sequence. For common spin-echo (SE) sequences, the signal equation for a single isochromat can be expressed as:
\begin{equation}
   m(\mathbf{r}) = K \rho (\mathbf{r})[1-e^{-\mathrm{TR/T_1}(\mathbf{r})}]e^{-\mathrm{TE/T_2}{(\mathbf{r})}},
   \label{eq:contrast1}
\end{equation}
where TR is the repetition time, TE is the echo time and $K$ is a constant. Manipulating TR and TE can lead to selective weightings of $\mathrm{T_1}$ and $\mathrm{T_2}$. For instance, moderate TR and short TE can produce $\mathrm{T_1}$-weighted contrast, which is useful in e.g. imaging fatty tissues with relatively low \Tone or white-, gray-matter distinction in brain imaging:
\begin{equation}
   m(\mathbf{r}) \approx K \rho (\mathbf{r})[1-e^{-\mathrm{TR}/\mathrm{T_1}(\mathbf{r})}].
   \label{eq:Tonew}
\end{equation}
Long TR and moderate TE can produce \Ttwon-weighted contrast, which can assist in e.g. detecting inflammation and edema: 
\begin{equation}
   m(\mathbf{r}) \approx K \rho (\mathbf{r})e^{-\mathrm{TE/T_2}{(\mathbf{r})}}.
   \label{eq:Ttwow}
\end{equation}
Meanwhile, long TR and short TE can produce PD-weighted contrast, which can be useful in e.g. assessment of joints:
\begin{equation}
   m(\mathbf{r}) \approx K \rho (\mathbf{r}).
   \label{eq:pdw}
\end{equation}
While SE sequences refocus magnetic field inhomogeneities through a refocusing pulse, gradient-echo (GRE) sequences lacking this mechanism are sensitive to additional signal decay governed by \Ttwosn, where $
\frac{1}{T_2^*} = \frac{1}{T_2} + \frac{1}{T_2'}$ with $T_2'$ denoting the portion of signal decay that can be recovered by a refocusing pulse, attributable to field inhomogeneities across a voxel. \Ttwosn-weighted contrast is more sensitive to local field variations, which can be beneficial for functional and susceptibility imaging.

Beyond manipulating TE and TR to gain sensitivity to \Tone and \Ttwo differences, diffusion of water molecules can also be leveraged to provide a unique contrast. Because moving spins experience different magnetic fields under the influence of a matched pair of gradients surrounding a refocusing pulse, they cannot be fully refocused at TE, thus attenuating the spin-echo signal under diffusion as \cite{jones2010diffusion},
\begin{equation}
   m(\mathbf{r}) = K \rho (\mathbf{r})[1-e^{-\mathrm{TR/T_1}(\mathbf{r})}]e^{-\mathrm{TE/T_2}{(\mathbf{r})}}e^{-bD(\mathbf{r})},
   \label{eq:diff}
\end{equation}
where $D(\mathbf{r})$ is the diffusion coefficient governing the attenuation, and $b=(\gamma G \Delta)^2(\Delta-\delta/3)$ depends on the amplitude ($G$), duration ($\delta$) and time separation ($\Delta$) of the gradients. By collecting multiple images with different $b$ values and diffusion directions, clinicians can assess the integrity of neural pathways. Note that MRI offers a diverse arsenal of contrasts, which go well beyond these common sequences (see \cite{Bernstein2004HandbookSequences} for further examples), to investigate the structure and function of tissues, making it an indispensable diagnostic tool. 

\subsection{Spatial Encoding in k-Space}
\begin{figure}[t]
\centering
\vspace{0.1cm}
\includegraphics[width=0.925\columnwidth]{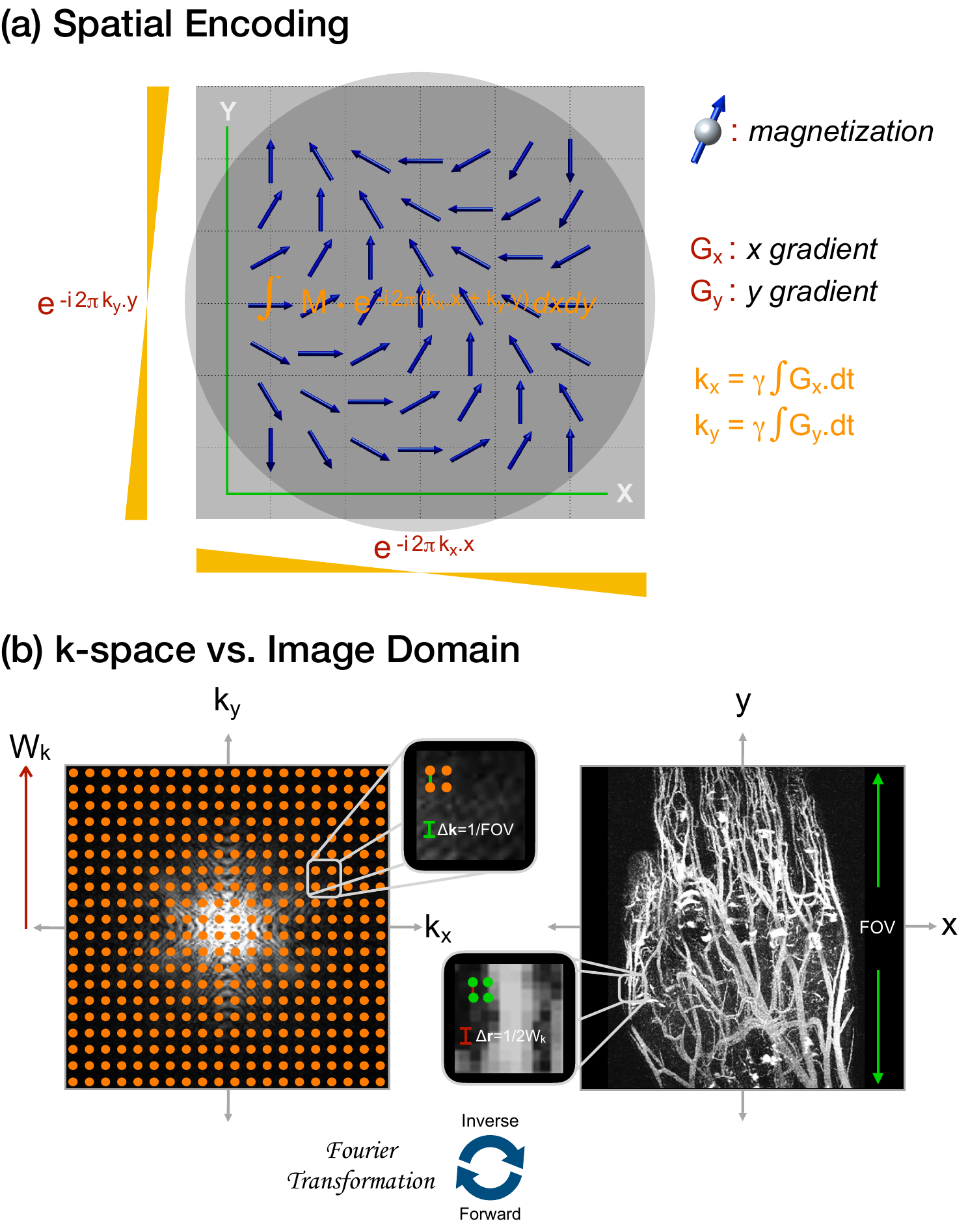}
\caption{\textbf{(a)} Illustration of spatial encoding via magnetic field gradients to localize signal origin. This encoding mechanism enables the systematic sampling of the tissue magnetization in the Fourier domain (i.e., k-space). \textbf{(b)} Visualization of the relationship between k-space where raw MRI data are acquired, and the image domain depicting the spatial distribution of tissue magnetization, linked through Fourier transformation.}
\label{fig:spatialencoding}
\end{figure}

Following an RF excitation, the pulse sequence performs spatial encoding to be able to resolve different spatial locations. Specifically, MRI data are acquired in \textit{k-space} by encoding spatial frequencies via magnetic field gradients that modulate the MR signal phase (Fig. \ref{fig:spatialencoding}a). For a gradient waveform $\mathbf{G}(t)=(G_x,G_y,G_z)$ that can be independently controlled across spatial axes, the shift in the phase of the MR signal is given as:
\begin{equation}
\label{eq:kspace}
    \Delta\phi = -2\pi \mathbf{k}(t) \cdot \mathbf{r} ;\,\, \mbox{such that }\, \mathbf{k}(t) = \frac{\gamma}{2\pi} \int \mathbf{G}(t) dt.
\end{equation}
In (\ref{eq:kspace}), $\mathbf{k}(t)=(k_x,k_y,k_z)$ denotes the evolution of the k-space trajectory across time, and $\gamma$ denotes the gyromagnetic ratio. The phase modulation from gradients causes the received signal $s(t)$ to reflect the Fourier transform of the underlying transverse magnetization $m(\mathbf{r},t)$: 
\begin{equation}
    s(t) = M(\mathbf{k})=\int_{\mathbb{R}^3} m(\mathbf{r},t) e^{-j 2\pi \mathbf{k}(t) \cdot \mathbf{r}} d\mathbf{r},
\end{equation}
Assuming that the magnetization is quasi-stationary across the k-space trajectory (i.e., $m(\mathbf{r},t)\approx m(\mathbf{r})$), its spatial distribution is hence related to ${M}(\mathbf{k})$ via an inverse Fourier transformation:
\begin{equation}
   m(\mathbf{r}) = \int_{\mathbb{R}^3} M(\mathbf{k}) e^{j 2\pi \mathbf{k} \cdot \mathbf{r}} d\mathbf{k}.
   \label{eq:IFT}
\end{equation}
Conventional data acquisition is performed discretely across a finite-extent Cartesian k-space grid, with sampling periods $\Delta\mathbf{k}=\Delta{(k_x,k_y,k_z)}$, and covered frequency ranges $\mathbf{W}_{\mathbf{k}}=W_{(k_{x},k_y,k_z)}$ across axes:
\begin{equation}
    \tilde{M}(\mathbf{k})=M(\mathbf{k}) \cdot \mathrm{III}(\mathbf{k} \varoslash \Delta \mathbf{k}) \cdot \mathrm{rect}(\mathbf{k} \varoslash \mathbf{W}_{\mathbf{k}}),
\end{equation}
where $\varoslash$ is Hadamard division, $\mathrm{III}$ is the Shah function, and rect is the rectangle function. Based on (\ref{eq:IFT}), an image $\tilde{m}(\mathbf{r})$ can be obtained as the inverse discrete Fourier transform of $\tilde{M}(\mathbf{k})$, related to the actual magnetization as (Fig. \ref{fig:spatialencoding}b): 
\begin{align}
   \tilde{m}(\mathbf{r}) = &m(\mathbf{r}) \circledast \,\mathrm{III}(\Delta \mathbf{k} \odot \mathbf{r}) \, \nonumber \\ &\circledast W_{k_x}W_{k_y}W_{k_z} \mathrm{sinc}(W_{k_x}x) \mathrm{sinc}(W_{k_y}y) \mathrm{sinc}(W_{k_z}z),
   \label{eq:alias}
\end{align}
where $\circledast$ is convolution, and $\odot$ is Hadamard product. Convolution with $\mathrm{III}(\cdot)$ causes $\tilde{m}(\mathbf{r})$ to contain replicas of ${m}(\mathbf{r})$, yielding an aliasing-free field-of-view (FOV) of size $1/\Delta \mathbf{k}$ when the Nyquist sampling criterion is met, i.e., the spatial extent of the object $\mathbf{W}_\mathbf{r}=W_{(x,y,z)}$ does not exceed $1/\Delta \mathbf{k}$. Convolution with sinc$(\cdot)$ terms sets the spatial resolution as $\Delta \mathbf{r} \approx 1/2\mathbf{W}_\mathbf{k}$. Note that (\ref{eq:alias}) describes 3D Fourier encoding in $x$-$y$-$z$ dimensions, though many MRI sequences can also use 2D encoding in $x$-$y$ with slice-selective excitation along $z$ \cite{Bernstein2004HandbookSequences}. For notational convenience, we denote the MR image obtained through discrete k-space sampling of the magnetization as $m$ in the remainder of the manuscript.

\subsection{Reconstruction: An Inverse Problem}
\begin{figure}
\centering
\vspace{0.1cm}
\includegraphics[width=\columnwidth]{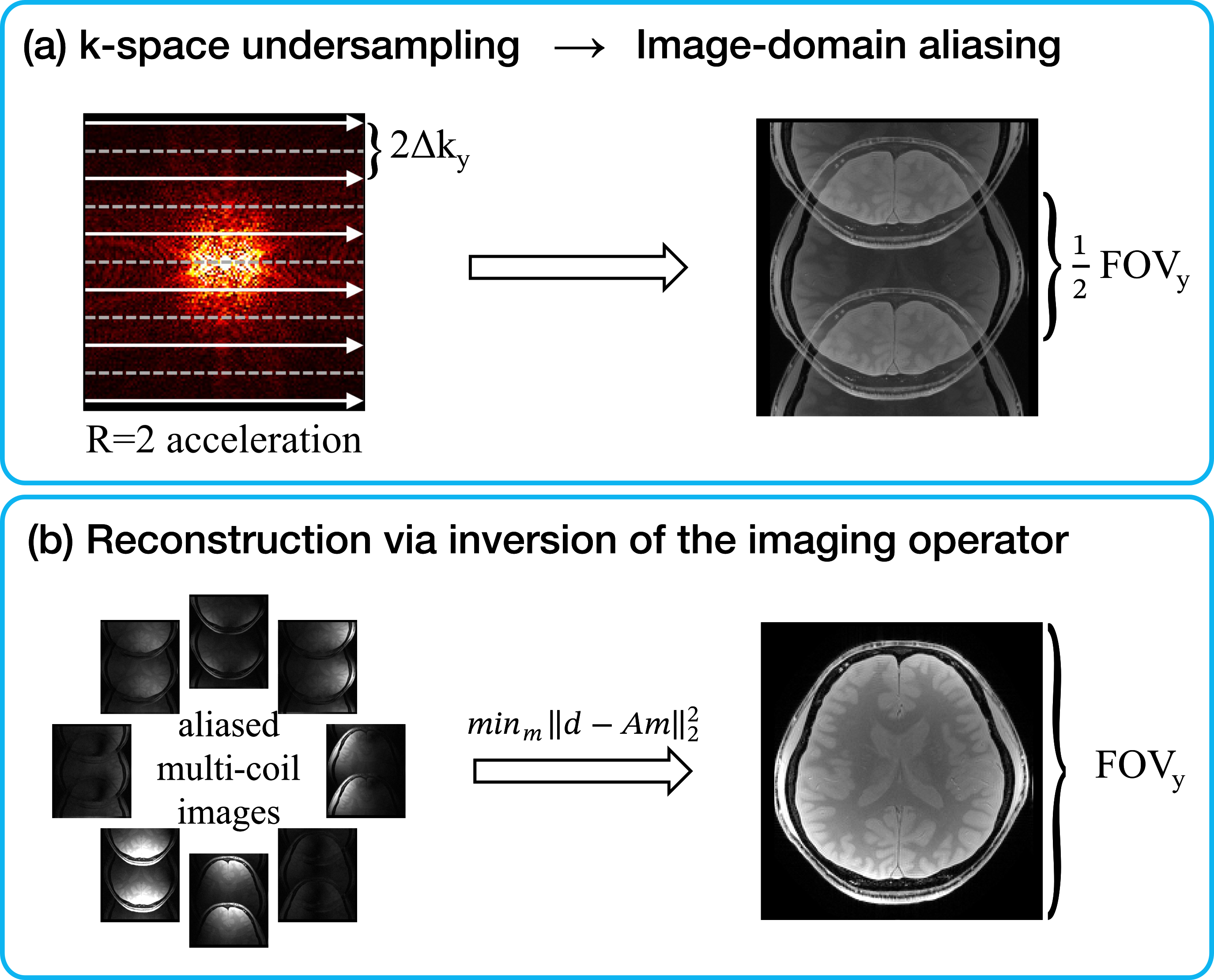}
\caption{\textbf{(a)} Undersampling in k-space leads to aliasing artifacts in the image domain following inverse Fourier transformation. \textbf{(b)} Multi-coil aliased images are processed to invert the imaging operator $A$, which encapsulates the effects of the sampling pattern, Fourier encoding, and coil sensitivities. A reliable inversion can produce an artifact-reduced, coil-combined image that approximates the fully-sampled reference.}
\label{fig:reconstruction}
\end{figure}

In an MRI acquisition with an acceleration factor of $R$, a $1/R$\textsuperscript{th} subset of samples on the k-space grid are collected, as defined via a binary sampling pattern $\mathcal{P}(\mathbf{k})$. Furthermore, multi-element coil arrays are used that induce spatial sensitivity profiles $C(\mathbf{r})$ that multiply the magnetization distribution. These influences can be captured via an imaging operator $A$, linking the magnetization distribution to acquired data $d(\mathbf{k})$: 
\begin{align}
   d(\mathbf{k}) &= \mathcal{P}(\mathbf{k}) \mathcal{F} \{C(\mathbf{r})m(\mathbf{r})\},\\
   d(\mathbf{k}) &= A \, m(\mathbf{r}). \label{eq:forward}
\end{align}
Image reconstruction, i.e., recovering $m(\mathbf{r})$ from $d(\mathbf{k})$, hence can be cast as an inverse problem aiming to minimize the discrepancy between acquired and reconstructed data, derived via a complex Euclidean norm since noise in MRI data is bivariate Gaussian (Fig. \ref{fig:reconstruction}):
\begin{equation}
    m^*=\underset{m}{\mathrm{arg min}} \, || d-Am||^2
    \label{eq:inverse},
\end{equation}
whose analytical solution is given by \revhl{$m^*=A^\dagger d = (A^H A)^{\dagger} A^H d$}, with $A^H$ denoting the Hermitian adjoint of $A$ \revhl{and $\dagger$ denoting Moore-Penrose pseudoinverse}. Since the Fourier transform is unitary (i.e., $\mathcal{F}^{-1}=\mathcal{F}^H$), $A$ becomes relatively well-conditioned when the k-space grid is fully sampled (i.e., $\mathcal{P}=\mathbf{1}$). Under Cartesian k-space sampling, the solution can then be efficiently approximated by applying a coil-by-coil inverse Fast Fourier Transform (FFT) to the acquired data, followed by coil combination using sensitivity maps, yielding $m^* \approx A^H d$ \cite{phasedarray}. In accelerated MRI, however, undersampling may render the problem ill-posed, necessitating the use of regularization to incorporate additional prior information:
\begin{equation}
    m^*=\underset{m}{\mathrm{arg min}} \, || d-Am||^2 + \lambda \Psi(m),
    \label{eq:inverse_prior}
\end{equation}
where the first term enforces consistency to acquired data, and $\Psi(m)$ weighted by $\lambda$ enforces prior information. In classical methods, $\Psi(m)$ has an explicit formulation based on hand-crafted regularizers that impose assumptions on the data. In DL methods, $\Psi(m)$ has an implicit formulation based on data-driven regularizers learned from a training dataset.

\section{Classical Reconstruction Methods}
\begin{figure*}
\centering
\includegraphics[width=0.9\textwidth]{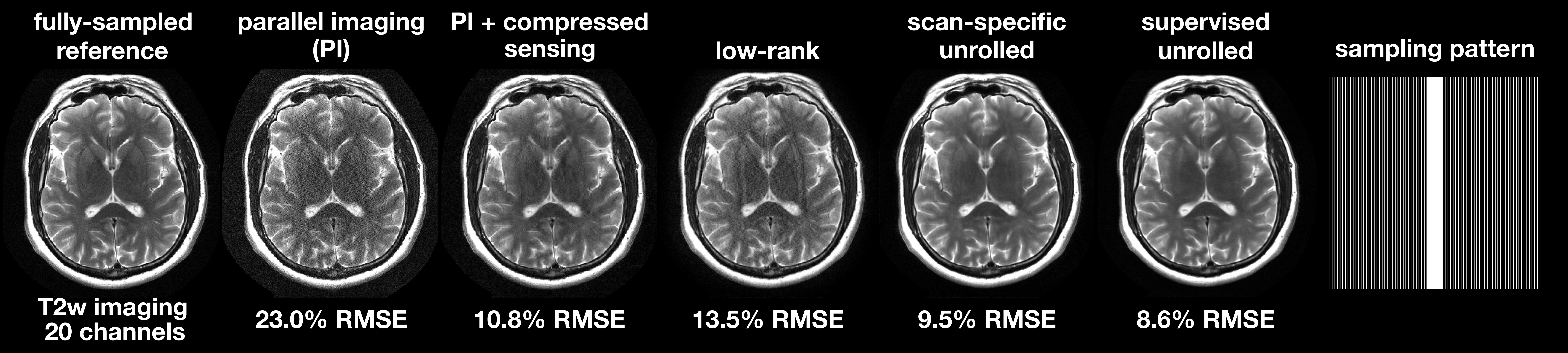}
\caption{Reconstruction of an axial T\textsubscript{2}-weighted brain image from undersampled k-space data using various methods. From left to right: fully-sampled reference, parallel imaging (PI), parallel imaging with compressed sensing (PICS), low-rank reconstruction, scan-specific unrolled network, supervised unrolled network, and the sampling pattern. Root-mean-squared error (RMSE) between reconstructed and reference images is reported below each method. PI and low-rank reconstructions show higher noise and residual aliasing, while PICS reduces noise but introduces block-like artifacts. DL methods achieve lower RMSE values, with visually reduced artifacts and improved sharpness relative to classical methods. The supervised model provides the strongest artifact suppression, albeit with a few subtle synthetic features, whereas the scan-specific model offers higher fidelity at the expense of residual artifacts.}
\label{fig:results}
\end{figure*}

Classical methods integrate hand-crafted physical priors (i.e., explicit signal models or assumptions on the data distribution) to regularize the solution of the inverse problem in (\ref{eq:inverse_prior}) (see Fig. \ref{fig:results} for example reconstructions). Below, we review major categories, including linear regression/interpolation, sparsity, low rank, and subspace models, each leveraging different structural properties of MRI data.

\textit{\color{nblue}\normalfont\fontsize{9}{11}\selectfont\everymath={\sf}\sf\itshape Linear Regression/Interpolation Methods:}
PI reduces the FOV deliberately by omitting acquisition of k-space lines, and attempts to resolve aliasing artifacts in image (e.g., SENSE \cite{SENSE}) or k-space domain (e.g., GRAPPA \cite{griswold2002generalized}). In image-domain techniques,
coil sensitivity profiles are used explicitly to unalias voxels by solving the linear regression problem in (\ref{eq:inverse_prior}). For the basic unregularized case ($\lambda = 0$), this problem admits a closed-form, non-iterative least-squares solution based on the Moore–Penrose pseudoinverse (i.e., \revhl{$m^*=A^{\dagger}d$}) \cite{SENSE}. \revhl{While this non-iterative solution is commonly used for the basic case, iterative solvers can provide improved numerical stability and efficiency in higher-dimensional or more complex reconstruction settings.} For instance, a gradient descent algorithm\footnote{An implementation of SENSE is available at \url{https://github.com/tutorial-MRI-recon/tutorial/tree/main/basic_sense_cs}.} can compute $m^*=\underset{m}{\mathrm{argmin}} \, f_{pi}(m)=\underset{m}{\mathrm{argmin}} \, ||d-Am||^2$: 
\begin{equation}
    m_{t+1}=m_t - \alpha \cdot \partial f_{pi}(m_t),
\end{equation}
where $\alpha$ is a step size that needs to be chosen appropriately, and $m_t$ and $m_{t+1}$ denote the image estimates at steps $t$ and $t+1$. Evaluating the gradient above leads to the update rule,
\begin{equation}
    m_{t+1}=m_t - \alpha \cdot A^H(A m_t - d),
\end{equation}
which is run until a convergence criterion is satisfied. 

Unlike SENSE, GRAPPA uses inter-coil relationships to interpolate missing k-space data. This interpolation is performed across coils and within a small k-space neighborhood:
\begin{equation}
M_j(k_x,k_y)=\sum_i \sum_{u \in U} \sum_{v \in V}g_j(i,\mu,v) M_i(k_x +u \Delta k_x , k_y + v \Delta k_y),
    \end{equation} 
with $i = 1, ..., N_c$ where $N_c$ is the number of coils. Here, $M_j(k_x,k_y)$ is the target missing k-space sample in the $j^{th}$ coil and $g_j$ is the k-space ``kernel'' used for interpolating the data for the $j^{th}$  coil. This kernel computes a linear combination of the acquired points $M_i(k_x +u \Delta k_x , k_y + v \Delta k_y)$ inside a small neighborhood in k-space, denoted with $U$ and $V$, across all coils. The kernel weights $g_j$ are estimated from low-resolution, fully-sampled calibration data, which is collected via a pre-scan or formed from the central portion of undersampled data.  

Note that the estimation of coil sensitivity profiles $C$ from calibration data is critical to construct an accurate estimate of the imaging operator in image-domain or analogously the interpolation kernel in k-space techniques. While heuristic approaches such as polynomial fitting can be used to estimate smoothly-varying profiles, structured methods have become prominent that estimate them as eigenvectors in the null space of a calibration matrix formed using calibration data (e.g., ESPIRiT \cite{uecker2014espirit}, PISCO\cite{PISCO}). Beyond such a priori estimation strategies, it is also possible to formulate nonlinear optimization problems that jointly estimate coil sensitivity profiles and the underlying image (e.g., JSENSE \cite{ying2007mrm}, IRGNM \cite{uecker2008mrm}). These nonlinear approaches can yield more accurate reconstructions, albeit at the cost of increased computational complexity.

\textit{\color{nblue}\normalfont\fontsize{9}{11}\selectfont\everymath={\sf}\sf\itshape Sparsity Methods:}
$\Psi(m)$ in (\ref{eq:inverse_prior}) can make use of a sparsity prior as popularized by CS methods that recover images from compressible measurements \cite{lustig2007sparse}. These priors are often formulated as $\Psi(m)= ||\Phi m||_1$, where $\Phi$ performs linear transformation to a domain (e.g. wavelet or gradient domain), whose coefficients are assumed to be compressible. Such sparsity priors can be incorporated into PI to perform a PICS reconstruction\footnote{An implementation of PICS is available at \url{https://github.com/tutorial-MRI-recon/tutorial/tree/main/basic_sense_cs}.} \cite{block2007undersampled,ying2009mrm,otazo2010combination}:
\begin{equation}
    m^*=\underset{m}{\mathrm{arg min}} \, f_{pics}(m) =\underset{m}{\mathrm{arg min}} \, || d-Am||^2+\lambda  ||\Phi m||_1.
    \label{eq:pics}
\end{equation}
When $\Phi$ is orthogonal or a tight frame, (\ref{eq:pics}) can be solved via a proximal gradient descent algorithm \cite{beck2009fast},
\begin{align}
v_t &= m_t - \alpha A^H(A m_t - d), \\
m_{t+1} &= \Phi^H \mathcal{S}_{\alpha \lambda}(\Phi v_t),
\end{align}
where $\alpha$ denotes step size, $v_t$ denotes an intermediate iterate, and $\mathcal{S}_{\alpha \lambda}(\cdot)$ denotes element-wise soft-thresholding with a threshold of $\alpha \lambda$. Note that, unlike unregularized PI methods, PICS algorithms can benefit from the noise-like aliasing interference elicited by randomized undersampling patterns \cite{murphy2012fast}.

\textit{\color{nblue}\normalfont\fontsize{9}{11}\selectfont\everymath={\sf}\sf\itshape Low-Rank Methods:}
Low-rank methods typically extract local patches from k-space data and stack them as rows of a structured matrix \cite{shin2014calibrationless,otazo2015low,LORAKS,mani2017multi,ALOHA}. Linear dependencies across coils, limited spatial support, phase smoothness, and transform-domain sparsity properties confer this matrix with low-rankness, which can be exploited as prior information  \cite{LORAKS,shin2014calibrationless,haldar2015autocalibrated,haldar2016p,kim2017loraks,bilgic2018improving}. Specifically, given $d$, local neighborhood of k-space samples are arranged into $\mathcal{H}(k)$, forming a Hankel or Toeplitz matrix that exhibits low-rank characteristics,
\begin{equation}
\text{rank}(\mathcal{H}(k)) \leq p, \quad \text{such that} \quad d = Am, 
\end{equation}
However, directly enforcing this rank constraint results in a non-convex, NP-hard problem, which is challenging to solve. To make the problem tractable, a typical approach is to convert the rank constraint into a minimization problem\footnote{An implementation of LORAKS is available at \url{https://github.com/tutorial-MRI-recon/tutorial/tree/main/low_rank}.},
\begin{equation}
m^* = \underset{m}{\mathrm{arg min}}  \| d - Am \|^2 + \lambda \Psi(\mathcal{H}(k)), 
\end{equation}
where $\Psi(\cdot)$ is a surrogate rank penalty. Commonly used formulations for $\Psi(\cdot)$ include \cite{LORAKS, haldar2015autocalibrated, haldar2016p, kim2017loraks},
\begin{equation}
\Psi(\mathbf{X}) = \min_{\mathbf{T}, \, \text{rank}(\mathbf{T}) \leq p} \| \mathbf{X} - \mathbf{T} \|_F^2 = \sum_{i>p} \sigma_i^2
\end{equation} 
or \cite{ALOHA,ongie2017fast},
\begin{equation}
\Psi(\mathbf{X}) = \| \mathbf{X} \|_*  = \sum_i \sigma_i
\end{equation}
where $\sigma_i$ denotes the singular values of matrix $\mathbf{X}$, and $p$ is a user-defined rank parameter.

\textit{\color{nblue}\normalfont\fontsize{9}{11}\selectfont\everymath={\sf}\sf\itshape Subspace Methods:}
Quantitative MRI (qMRI) reconstructions can leverage the knowledge of spin evolution by creating a low-rank subspace $\Gamma$ for MR signals, where $\Gamma$ is formed with singular value decomposition (SVD) of a dictionary generated through Bloch signal simulations or learned from sample image patches \cite{Ravishankar2011a}. Subspace coefficients $s$ in $\Gamma$ can be estimated from undersampled k-space data, such that $\Gamma s$ is an accurate approximation to the measured signal evolution  \cite{tamir2017t2}. This estimation can be performed as, 
\begin{equation}
    s^* =\underset{s}{\mathrm{arg min}} \, || d-A\Gamma s||^2+\lambda \cdot 
    \Psi(s) 
    \label{eq:subspace}
\end{equation}
where the regularizer $\Psi(s)$ could impose sparsity or low-rank constraints on the subspace coefficients. Having obtained the coefficients, images with various different contrasts can then be synthesized. A similar approach is adopted in MR Fingerprinting (MRF) \cite{ma2013magnetic}, where Bloch simulations lend themselves to creation of a dictionary that explains the behavior of transverse signal as the flip angle and TRs are evolving throughout the acquisition. Rather than using such a dictionary only at the signal matching step after image reconstruction, it can be incorporated into the reconstruction process itself, so that dictionary coefficients that approximate the acquired data in a parsimonious manner can be directly estimated \cite{zhao2018improved}. This subspace approach can help mitigate relaxation-related blurring during the qMRI echo train, and boost SNR and mitigate artifacts in MRF \cite{cao2022optimized}.

Note that the subspace formulation can be viewed as a linear approximation of the underlying MR signal evolution. Rather than projecting signals into a fixed low-rank basis, an alternative is to explicitly enforce the nonlinear signal equations during reconstruction, enabling direct estimation of quantitative parameters (e.g., relaxation times) from undersampled data \cite{sumpf2011jmri}. Such nonlinear formulations can improve parameter accuracy by adhering more closely to MR physics, but they lead to nonlinear inverse problems that are often computationally more demanding to solve. These nonlinear formulations are applicable to steady-state sequences that admit a closed-form expression for their signal evolution, such as those in (\ref{eq:contrast1}-\ref{eq:Ttwow}).

\section{Deep-Learning Reconstruction Methods}
DL methods learn data-driven priors from exemplar MRI datasets to help solve (\ref{eq:inverse_prior}). To structure discussions on DL, we dissect methods in the literature along three fundamental axes: modeling framework, network architecture, and learning paradigm. This tripartite view reflects the core design choices that govern how DL reconstruction methods incorporate domain knowledge, compose their networks, and learn from data.

\begin{figure*}[t]
  \centering
  \includegraphics[width=0.805\textwidth]{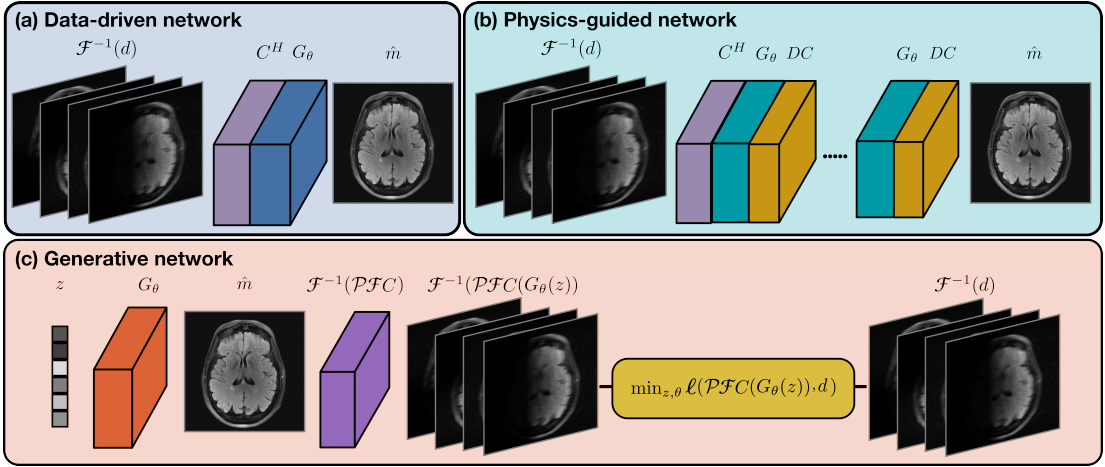}
  \caption{\textbf{(a)} A data-driven network $G_{\theta}$ reconstructs an MR image $\hat{m}$ from a linear transformation of undersampled multi-coil data $d$. In this example, the linear transformation is a zero-filled Fourier reconstruction ${m}_u=A^{H}d$, obtained by inverse Fourier transformation of $d$ followed by coil combination based on estimated profiles $C$. The reconstruction $\hat{m}$ is obtained via a forward pass of $m_u$ through network layers. \textbf{(b)} A physics-guided network reconstructs an MR image from a linear transformation of undersampled data. In this example, the zero-filled Fourier reconstruction $m_u$ is forward passed iteratively through interleaved network layers $G_{\theta}$ and data-consistency (DC) layers. In DC layers, the input coil-combined image is first projected onto a multi-coil image based on $C$, data consistency is enforced in k-space, and the resultant multi-coil image is coil-combined. \textbf{(c)} A generative network $G_{\theta}$ trained to capture the distribution of high-quality MR images is combined with the imaging operator to reconstruct an MR image at inference. In this example, $G_{\theta}$ synthesizes a random MR image from a latent representation $z$—this may be a random code (GANs), Gaussian noise (diffusion), or a latent code produced by an encoder. $G_{\theta}$ and $z$ can be jointly optimized to ensure that reconstructed k-space data are consistent with acquired data $d$ based on the imaging operator $A$.}
  \label{fig:dnn_methods}
\end{figure*}

\subsection{Modeling Frameworks}
To avoid overlapping terminology and blurred boundaries that may obscure complementary strengths and limitations, here we categorize current DL methods under three distinct frameworks: data-driven, physics-guided, and generative (Fig. \ref{fig:dnn_methods}; see Fig. \ref{fig:results} for example reconstructions). This taxonomy is grounded in fundamental distinctions regarding how DL methods leverage domain knowledge, i.e., how they incorporate data-driven and physical priors, to solve the inverse problem. Accordingly, data-driven and physics-guided models refer to conditional networks\footnote{Conditional networks generate outputs based on a measured input and a learned mapping, effectively modeling the conditional distribution $P(\text{output} \mid \text{input})$.} trained to predict MR images that would resemble reconstructions of fully-sampled data, using as input a linear transformation of undersampled data, and differing in whether they exploit physical priors. By contrast, generative models refer to unconditional networks\footnote{Unconditional networks generate outputs without an explicit measured input, often leveraging random latent variables, and model the marginal distribution $P(\text{output})$.} trained to synthesize random MR images that are not tied to undersampled data, integrating physical priors only during inference. This classification partly diverges from the broader machine learning convention, where `generative models' can encompass any architecture capable of producing stochastic images. In MRI reconstruction, however, we argue that the critical distinction lies not in output stochasticity, but in how models fundamentally approach the inverse problem. Conditional networks—even those with stochastic components like adversarial or variational architectures—receive undersampled data as explicit input and are thus more appropriately classified within our data-driven or physics-guided categories based on their treatment of physical priors. 

\subsubsection{Data-Driven Models} 
Instead of relying on predefined priors, these methods learn data-driven priors directly from available datasets \cite{Yu2018c,Mardani2019b,nath2020datadriven,wang2022b}. In a common formulation, a network is trained to map from undersampled to fully-sampled data, using a training dataset $\left\{ \left(d^{\revhl{\mathrm{tr}}(j)}, m^{\revhl{\mathrm{tr}}(j)}\right) \right\}_{j=1}^{N_{\revhl{\mathrm{tr}}}}$ where $j$ is the sample index, $N_{\revhl{\mathrm{tr}}}$ is the number of training samples, $d^{\revhl{\mathrm{tr}}(j)}$ denotes acquired data for the $jth$ sample, and $m^{\revhl{\mathrm{tr}}(j)}$ is the ground-truth image. Note that the network input/output can be expressed in image domain, k-space or both, with image-domain networks often using the least-squares solution $m_u=A^Hd$ (i.e., zero-filled Fourier reconstruction) as input, and trying to recover the ground-truth image $m$. The training objective for a data-driven (dd) network can then be expressed as:
\begin{equation}
\mathcal{L}_{\text{dd}}(\theta) = \mathbb{E}_{(d^{\revhl{\mathrm{tr}}}, m^{\revhl{\mathrm{tr}}})} \left[ \ell \big(G_\theta(m_u^{\revhl{\mathrm{tr}}}), \, m^{\revhl{\mathrm{tr}}}\big) \right]
\label{eq:net_training}
\end{equation}
where $\mathbb{E}$ denotes expectation, $G_\theta$ is the network parameterized by $\theta$, and $\ell(\cdot,\cdot)$ denotes a loss function that reflects the discrepancy between network-predicted and ground-truth reconstructions. Note that this discrepancy can be measured in image domain as depicted in (\ref{eq:net_training}), in k-space, or both. Furthermore, while the $\ell_2$-norm is a common choice for comparing network and ground-truth reconstructions—mirroring its role in quantifying data consistency—alternative losses are often employed to promote desired image attributes. Prominent examples include $\ell_1$, hybrid $\ell_1$–$\ell_2$, perceptual, adversarial, or diffusion losses \cite{goodfellow2016deep}.

During inference, the trained network with optimized parameters $\theta^*$ processes undersampled data $d$ to predict $\hat{m}$ through a forward pass across the network architecture:
\begin{equation}
\hat{m} = G_{\theta^*}(m_u).
\end{equation}

Learning an end-to-end mapping from undersampled to fully-sampled data, purely data-driven models are computationally efficient and can outperform traditional methods. Yet, they may suffer performance losses due to the lack of an explicit mechanism to enforce fidelity to the imaging operator or acquired data \cite{radhakrishna2023jointly}. As remedy, some approaches embed data-driven networks—typically pretrained for denoising or artifact suppression—within iterative optimization algorithms \cite{ahmad2020plug,liu2020rare}. This keeps the network agnostic to the imaging operator during training, while imposing physical priors externally at inference—a middle ground between purely data-driven and tightly integrated physics-guided designs.

\subsubsection{Physics-Guided Models} To benefit from knowledge of acquired k-space locations and coil sensitivities, physics-guided models typically inject data-consistency blocks based on the imaging operator within the network architecture \cite{hammernik2022physicsdriven,MoDl,hammernik2018learning,Zhou_2020_CVPR,polak2020joint,sun2019deep,yaman2020, Ryu2021-oy,liu2021regularization,ramzi2022nc,Lam2023}. A common motivation for this design is a composite objective that balances physics-driven data consistency with data-driven regularization:
\begin{equation}
m^* = \underset{m}{\mathrm{arg min}} \, \|d - Am \|^2 + \lambda \, \ell (G_{\theta}(m),m),
\label{eq:unroll_obj}
\end{equation}
where $G_{\theta}$ denotes the network module. In practice, unrolled models do not directly solve (\ref{eq:unroll_obj}), but instead alternate a learned prior block with a quadratic data-consistency update as described in (\ref{eq:unroll_reg}-\ref{eq:unroll_DC}). This iterative structure resembles fixed-point iterations, and has close connections to proximal splitting methods \cite{combettes2011proximal, chan2016plug, romano2017little, ryu2019plug}:
\begin{align}
z^{(i+1)} &= G_{\theta}(m^{(i)}), \label{eq:unroll_reg} \\
m^{(i+1)} &= \underset{m}{\mathrm{arg min}} \, \|d - Am\|^2 + \lambda \, \ell (z^{(i+1)},m), \label{eq:unroll_DC} 
\end{align}
where $I$ is the total number of iterations, and $0 \leq i<I$. The training objective for a physics-guided (pg) network can be formulated as:
\begin{equation}
\mathcal{L}_{\text{pg}}(\theta) = \mathbb{E}_{(d^{\revhl{\mathrm{tr}}}, m^{\revhl{\mathrm{tr}}})} \big\{\ell(m^{(I)},m^{\revhl{\mathrm{tr}}})\big\}
\end{equation}
Image-domain unrolled networks typically set the input as $m^{(0)}=m_u$, and given a trained network, the reconstructed image during inference is taken as $\hat{m}=m^{(I)}$\footnote{An implementation of an unrolled model based on hybrid $\ell_1$-$\ell_2$ loss is available at \url{https://github.com/tutorial-MRI-recon/tutorial/tree/main/supervised_dl}.}.

Traditional methods for quantitative MRI often rely on analytical models that describe how tissue parameters affect the measured signals based on the physics of spin dynamics (e.g., mono-exponential decay in multi-echo spin-echo sequences for \Ttwo mapping). Recent physics-guided DL methods also integrate such analytical priors to estimate quantitative maps of tissue parameters $q$ from undersampled data \cite{liu2019mantis,Mani2020Model-BasedMRI}:
\begin{equation}
    q^* =\underset{q}{\mathrm{arg min}} \, || d-A h(q)||^2+\lambda \, \ell(G_{\theta}(q),q),
    \label{eq:signal_aug}
\end{equation}
where $h(\cdot)$ is the analytical signal model. By jointly optimizing over parameter and data fidelity, such unrolled networks can mitigate errors that arise from treating image reconstruction and parameter mapping disjointly. 

Physics-guided models incorporate the imaging operator to improve reconstruction performance, but they may face generalization issues when tested out-of-domain—e.g. with different undersampling patterns or rates—and become computationally expensive when using a large number of cascades.

\begin{figure*}[t]
  \centering
  \includegraphics[width=0.855\textwidth]{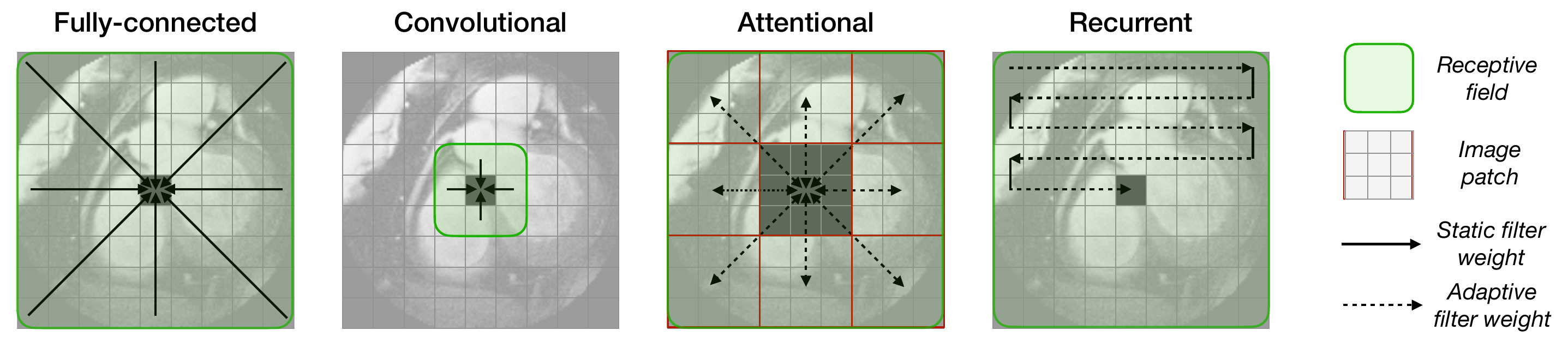}
  \caption{Receptive fields for fully-connected, convolutional, attentional, and recurrent (e.g., state-space) operators, illustrating differences in filtering primitives. Fully-connected operators perform static global filtering across the image. Convolutional operators perform static local filtering constrained to a compact neighborhood. Attentional operators perform input-adaptive global filtering by computing pair-wise similarities of image tokens (e.g., patches). Recurrent operators perform input-adaptive global filtering by aggregating high-order dependencies across image pixels through sequence modeling.}
  \label{fig:operators}
\end{figure*}

\subsubsection{Generative Models} 
\label{sec:gen_prior}
Unlike models subject to task-specific training involving undersampled and fully-sampled data, generative priors agnostic to the imaging operator (i.e., the k-space sampling patterns and coil sensitivity maps) have come forth as a promising alternative to boost generalization by learning the distribution of fully-sampled data alone \cite{quan2018compressed,tezcan2018mr}. For inference, a generative prior that synthesizes MR images can be combined with the imaging operator, typically through a conventional optimization problem. A variety of machine learning approaches from autoencoding to flow-matching models are available for building generative priors  \cite{raut2024generativeaivisionsurvey}, yet arguably adversarial and diffusion models have gained particularly widespread adoption.  

Earlier works utilized Generative Adversarial Networks (GANs) \cite{narnhofer2019inverse,korkmaz2022unsupervised} comprising a generator network $G_{\theta}$ that synthesizes images from latent variables, and a discriminator network $D_{\theta}$ that distinguishes between actual and synthetic images, trained via the following objective:
\begin{align}
\mathcal{L}_{\text{GAN}}(\theta)= & \mathbb{E}_{m \sim m^{\revhl{\mathrm{tr}}}} \left[ \log D_{\theta}(m) \right]  \nonumber \\ 
   & + \mathbb{E}_{z \sim p_z(z)} \left[ \log \left( 1 - D_{\theta}(G_{\theta}(z)) \right) \right],
\end{align}
where $p_z(z)$ denotes the distribution of latent variables. During inference, the latent code and, sometimes, the generator itself are optimized so as to satisfy fidelity with the imaging operator and acquired data:
\begin{equation}
(z, \theta)^* = \underset{z, \theta}{\mathrm{argmin}} \, \ell (AG_{\theta}(z),d) 
\label{eq:gan_update}
\end{equation}
where $z$ denotes a vector of random latent variables, and $\ell$ is a loss function that measures discrepancy between reconstructed and acquired k-space data (e.g., $\ell_2$-norm, $\ell_1$-norm, or hybrid $\ell_1$-$\ell_2$ norm). The final reconstruction can be computed via the generator network as:
\begin{equation}
\hat{m} = G_{\theta^{*}}(z^{*})
\end{equation}

More recent approaches have leveraged diffusion-based priors that operate through two complementary processes: a forward process that progressively adds noise to training images, and a reverse process that learns to denoise images step-by-step \cite{jalaln2021nips,chung2022media,uecker2023mrm}. During training, the forward process systematically corrupts clean images by adding Gaussian noise over $T$ timesteps, while the reverse process trains a neural network $G_{\theta}$ to predict and remove this added noise. The training objective can be expressed as:
\begin{align}
\mathcal{L}_\text{diff}(\theta) = \mathbb{E}_{t \sim [1, T], m_0^{\revhl{\mathrm{tr}}}, \epsilon} & \left[ \| \epsilon - \right. \nonumber  \\
       &  \left.  G_{\theta}\left( \sqrt{\bar{\alpha}_t} m_0^{\revhl{\mathrm{tr}}} + \sqrt{1 - \bar{\alpha}_t} \epsilon, t \right) \|^2 \right],
\end{align}
where $m_0^{\revhl{\mathrm{tr}}}$ refers to ground-truth images from the training set, \( \epsilon \sim \mathcal{N}(0, I) \) is Gaussian noise, \( G_{\theta}(\cdot, t) \) is the neural network that predicts the added noise at timestep \( t \), and \( \bar{\alpha}_t = \prod_{i=1}^{t} \alpha_i \) defines the cumulative noise schedule that controls the amount of noise added during the forward process. During inference, only the reverse process is used to reconstruct the MR image corresponding to an undersampled acquisition, starting from a pure Gaussian noise image. To do this, the denoising process is guided between successive steps to ensure that the synthesized images remain faithful to acquired data \cite{ssdiffrecon}, analogously to an unrolled network operating with $T$ iterations. At each step $t$, the sampling procedure for a diffusion model can be expressed as:
\begin{align}
\tilde{m}_{t-1} &= G_{\theta}(m_t,t), t \in [T, \, 0), \\
m_{t-1} &=  \tilde{m}_{t-1} + A^H ( d - A( \tilde{m}_{t-1} ) ),
\end{align}
where $\tilde{m}_{t-1}$ is the image at step $(t-1)$ that can be predicted using $G_{\theta}$, $m_{t-1}$ is the corrected data-consistent version, and $m_T=\epsilon$ represents the starting point of pure Gaussian noise. The final reconstructed image is taken as $\hat{m}=m_{0}$. Image quality can be further enhanced by adapting the generative prior itself to acquired data, as in (\ref{eq:gan_update}) \cite{gungor2022adaptive}.

Generative methods decouple the imaging operator from the training of the network to arrive at a formulation that improves generalization. However, they often suffer from prolonged inference times, and determining appropriate stopping criteria for test-time adaptation can be challenging.

\subsection{Network Architectures}
Although the modeling framework dictates the overarching network structure for reconstruction models, a range of architectural decisions from operator type to contrast configuration must be made when composing individual layers. 

\subsubsection{Operator Type} Each layer applies mathematical operators that capture spatial dependencies in the input--in conjunction with normalization and non-linear activation functions-- enabling the extraction of task-relevant features that are propagated through the network (Fig. \ref{fig:operators}). These operators define distinct filtering primitives, imparting varying degrees of representational power and computational efficiency.

\textbf{Fully-connected operators} establish an explicit relationship between each pair of elements in the input feature map versus the output feature map:
\begin{align}
m_\text{out}[j] = \sum_{i} w[i,j], m_\text{in}[i],
\end{align}
where $m_\text{in} \in \mathbb{C}^M$ denotes the vectorized input map, $m_\text{out} \in \mathbb{C}^M$ is the output map, and $w \in \mathbb{C}^{M \times M}$ is the static filter weights of the operator. As complex-valued operators are relatively uncommon in DL programming libraries, practical implementations often map real/imaginary or magnitude/phase components across the channel dimension of feature maps and prescribe real-valued operators \cite{Dar2017}. Commonly found in implicit generative methods that synthesize images as a function of spatial coordinates \cite{liang2025analysis}, fully-connected operators enable modeling of all pairwise interactions between input and output elements \cite{Kwon2017}, yet they are parameter-intensive and often computationally prohibitive for many network designs.

\textbf{Convolutional operators} use a translation-invariant kernel to filter the input feature map: 
\begin{align}
    m_\text{out}[u,v] = \sum_{i} \sum_{j} w[u-i, v-j] m_\text{in}[i,j],
\end{align}
where $m_{\text{in},\,\text{out}} \in \mathbb{C}^{U \times V}$, and $w \in \mathbb{C}^{r \times r}$ denotes the static filter weights with $r \ll U,V$. Convolutional operators are parameter-efficient due to their localized receptive fields and hence easier to train on compact datasets \cite{RAKI,MoDl}, which have contributed significantly to their widespread use in DL models for MRI reconstruction. However, these benefits come at the cost of a bias toward localized dependencies between neighboring elements, which limits sensitivity to long-range interactions across the input map.

\textbf{Attentional operators} compute interactions across the input feature map \( m_\text{in} \in \mathbb{C}^{U \times V} \) by partitioning it into \( N = \frac{UV}{P^2} \) patches \( \{m_\text{in}^k\}_{k=1}^N \), each of size \( P \times P\). Each patch is flattened and linearly projected into a vector \( z_k \in \mathbb{C}^D \). The output patch representations \( \{z'_j\}_{j=1}^N \) are derived as \cite{dosovitskiyViT,korkmaz2022unsupervised}:  
\begin{align}
    z'_j &= \sum_{k=1}^N \alpha_{j,k} (z_k W_V), \\
    \alpha_{j,k} &= \frac{\exp\left( (z_j W_Q)(z_k W_K)^\top / \sqrt{D} \right)}{\sum_{l=1}^N \exp\left( (z_j W_Q)(z_l W_K)^\top / \sqrt{D} \right)},
\end{align}  
where \( W_Q, W_K, W_V \in \mathbb{C}^{D \times D} \) are learnable query, key, and value projection matrices, and the attentional filter weights $\alpha$ are input adaptive. The output feature map \( m_\text{out} \in \mathbb{C}^{U \times V} \) is reconstructed by reshaping and aggregating the output patches \( \{z'_j\} \). Attention enables modeling of long-range interactions across the input map, but it incurs higher computational and memory costs compared to convolutional operators, particularly for high-resolution images with many patches \cite{guo2022reconformer,huang2022swin,feng2021task}.

\textbf{Recurrent operators} propagate information sequentially across the spatial dimensions, providing an alternative to attention primitives for capturing global context \cite{qin2018convolutional,Hosseini2020b,guo2021over}. A representative example is the state-space model (SSM) operator that describes the dynamics of hidden states \( h[i] \in \mathbb{C}^N \) in response to an input sequence of pixels \( m_\text{in}[i] \in \mathbb{C}^M \) indexed by \( i\) through a linear system \cite{gu2023mamba}:
\begin{align}
    h[i+1] &= A_\text{ssm} \, h[i] + B_\text{ssm} \, m_\text{in}[i], \\
    m_\text{out}[i] &= C_\text{ssm} \, h[i] + D_\text{ssm} \, m_\text{in}[i],
\end{align}
where \( A_\text{ssm} \in \mathbb{C}^{N \times N} \), \( B_\text{ssm} \in \mathbb{C}^{N \times M} \), \( C_\text{ssm} \in \mathbb{C}^{M \times N} \), \( D_\text{ssm} \in \mathbb{C}^{M \times M} \) are learnable matrices for adaptive filtering. Here, flattening the input feature map \( m_\text{in} \) into a 1D sequence via a specific pixel ordering across the spatial grid constrains the inherently multidimensional spatial relationships to a single traversal path. Yet, this sequentialization enables SSMs to summarize long-range spatial dependencies at lower computational cost compared to attentional operators \cite{huang2025enhancing,Mambarecon,Mambaroll}. Ongoing works explore alternative sequentialization approaches (e.g., raster scan, spiral, space-filling curves) to balance fidelity to image structure with efficiency \cite{gu2023mamba,I2IMamba}.

\subsubsection{Input Domain} 
Architectures vary in the domain in which they operate on MRI data, with each domain offering distinct advantages for capturing structural, frequency-specific, or semantic features essential to the reconstruction task.

\textbf{Image-domain networks} typically operate on aliased images, obtained via zero-filled inverse Fourier transformation of undersampled data \cite{hammernik2018learning}. Within physics-guided or generative frameworks, image-domain feature maps are often propagated between the image and Fourier domains to enforce data consistency based on the imaging operator \cite{MoDl}.

\textbf{Fourier-domain networks} directly process undersampled k-space data, offering an intrinsic advantage for enforcing data-consistency, as the data remain in the Fourier domain \cite{zhu2018image,kspaceDLMRI,fdb}. However, special care is often required when handling k-space inputs, as the energy distribution varies significantly across spatial frequencies.

\textbf{Hybrid-domain networks} employ separate branches or interleaved cascades to process both image- and Fourier-domain representations \cite{KikiNet,Wang2019,Zhou_2020_CVPR}. By combining spatially localized image features with globally distributed spectral features, these architectures can achieve enhanced reconstruction performance compared to single-domain counterparts.

\textbf{Latent-domain networks} abstract away from image or k-space representations by operating in a learned latent space. Such networks encode MRI data onto a lower-dimensional semantic representation, and decode the processed semantic representation back onto MRI data \cite{tezcan2018mr,Biswas2019,Liu2020HighlyPriors,FedGAT}. This strategy enables the more efficient processing of high-dimensional and multi-contrast MRI datasets.

\begin{figure*}[t]
  \centering
  \vspace{-0.1cm}
  \includegraphics[width=0.85\textwidth]{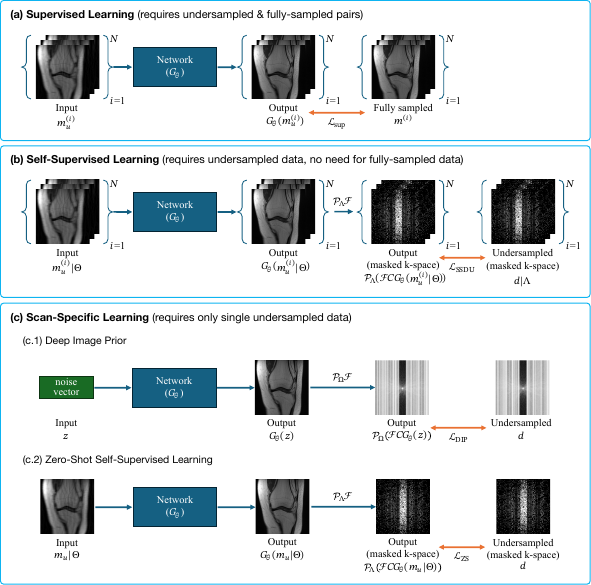}
  \caption{Different learning paradigms depending on availability and structure of training data. \textbf{(a)} Supervised learning performs population-level training of a reconstruction network using paired sets of undersampled and fully-sampled data, where training loss is expressed using ground-truth images derived from fully-sampled data. \textbf{(b)} Self-supervised learning performs population-level training using only undersampled data, in cases where it is not feasible to collect fully-sampled acquisitions. Training loss is expressed using a masked portion of undersampled data. \textbf{(c)} Scan-specific learning performs test-time training using undersampled data from an individual test subject. Reconstruction can be achieved via either an unconditional network that maps a noise vector $z$ onto an image as in deep image prior (DIP) methods, or a conditional network that maps a zero-filled Fourier reconstruction of undersampled data $m_u$ onto an image as in zero-shot (ZS) methods. In both cases, training loss is expressed using single-scan undersampled data.}
  \label{fig:dl_scheme}
\end{figure*}

\subsubsection{Contrast Configuration} Architectures can also differ in whether they process different tissue contrasts in an MRI protocol independently or jointly. This typically shapes the branching strategies within the network and directly impacts the network’s performance.

\textbf{Independent reconstruction} refers to the most common formulation adopted in MRI, where each contrast (e.g., T\textsubscript{1}, T\textsubscript{2}) is reconstructed separately using contrast-specific inputs \cite{wang2024review}. This approach simplifies modeling, and generally lowers model complexity. However, it also ignores potential redundancies and synergies across contrasts, which may limit reconstruction quality towards higher acceleration factors.

\textbf{Joint reconstruction} aims to reconstruct multiple contrasts acquired from the same subject simultaneously by leveraging shared structural features arising from the underlying common anatomy \cite{gong2015promise,bilgic2011multi}. Such networks typically receive multi-contrast inputs, either as multi-channel tensors or separate inputs fused through cross-modal attention or shared encoding \cite{sun2019deep,xiang2019rec,Do2020-xp,Ryu2021-oy}. Joint reconstruction can significantly improve performance, particularly under high accelerations, albeit requires careful registration of contrasts to avoid artifacts due to spatial misalignment and feature leakage \cite{Kopanoglu2020-iq}.

\textbf{Prior-conditioned reconstruction} refers to setups where the reconstruction of a target contrast is guided by one or more auxiliary contrasts \cite{dar2020prior,Kim2018-qd}. This formulation is particularly relevant in settings where one contrast carries a fully-sampled acquisition, and another has a heavily undersampled acquisition. Prior-conditioning mechanisms may include concatenation, attention, or learned cross-contrast priors \cite{liu2020multi}. Prior-conditioned reconstruction offers similar benefits to joint reconstruction, albeit proper alignment between contrasts is often key to model performance \cite{rizzuti2022joint}. Note that prior-conditioning does not have to be restricted to distinct contrasts; it can also be employed to process repeated scans of a given contrast. For instance, patient-specific conditioning can be realized to improve reconstruction performance using prior scans of the same contrast for a subject, even when acquired on different MRI systems (e.g., high-field versus low-field) \cite{oved2025deeplearningpersonalizedpriors}.

\subsection{Learning Paradigms}

The choice of training strategy for DL methods is closely tied to the availability and structure of training data. Depending on whether fully-sampled acquisitions are available to serve as ground truth, or whether a training set exists at all, different paradigms become suitable. Below, we review the three main paradigms commonly adopted in the literature that aim for population-level, scan-specific or hybrid training.

\subsubsection{Population-Level Training}
Population-level training aims to learn a reconstruction network from a cohort of training subjects, assuming access to multiple independent acquisitions. Depending on whether paired (fully-sampled and undersampled) data are available, supervised or self-supervised learning might be adopted.

\textbf{Supervised learning} assumes access to paired datasets $\{(m_u^{(i)}, m^{(i)})\}_{i=1}^N$, where $i$ is sample index, $N$ is the number of samples in the training set, $m_u$ is the zero-filled Fourier reconstruction of an undersampled acquisition (or its corresponding k-space data), $m$ is the ground-truth image Fourier reconstructed from a fully-sampled acquisition. The reconstruction network is trained to minimize the discrepancy between its prediction $\hat{m}^{(i)} = G_\theta(m_u^{(i)})$ and the ground truth (Fig. \ref{fig:dl_scheme}a).  general objective can be expressed as:
\begin{equation}
\mathcal{L}_{\text{sup}}(\theta) = \mathbb{E}_{(m_u, m)} \left[ \ell \left(G_\theta(m_u), m \right) \right],
\end{equation}
where $\mathbb{E}$ denotes expectation typically estimated via Monte Carlo sampling over training samples. Supervised methods yield high reconstruction accuracy when the distributions of training versus test samples are reasonably well aligned. However, they require large datasets of fully-sampled acquisitions, which are costly to collect in clinical settings \cite{knoll2020deep,heckel2024deep}.

Note that generative networks are typically trained to synthesize MR images derived from fully-sampled data, without considering the imaging operator. At inference time, they are combined with an inverse sampling technique (see Section \ref{sec:gen_prior}) to reconstruct undersampled data from test subjects. Formally, this setup aligns with the definition of unsupervised learning, as networks are trained without paired input-output examples, learning solely from the target distribution \cite{korkmaz2022unsupervised}.

\textbf{Self-supervised learning.} With MRI protocols that require significantly high spatial-temporal resolution and coverage, it may be infeasible to perform fully-sampled acquisitions to curate training sets in clinically acceptable scan times \cite{heckel2024deep,wang2024review}. In such scenarios, reconstruction models can be trained solely on undersampled acquisitions by adopting self-supervised learning \cite{Tamir2019,blumenthal2024mrm}. A common strategy partitions available measurements into training and supervision sets \cite{yaman2020,cole2021fast,liu2020rare}, where acquired k-space indices $\Omega$ are split into two disjoint sets $\Theta$ and $\Lambda$ such that $\Theta \cup \Lambda = \Omega$. The network is trained to reconstruct the signal over $\Lambda$ using predictions based on data from $\Theta$, as inspired by masked modeling techniques in computer vision \cite{dosovitskiyViT} (Fig. \ref{fig:dl_scheme}b):
\begin{equation}
\mathcal{L}_{\text{SSDU}}(\theta) = \mathbb{E}_{m_u} \left[ \ell \left( \mathcal{P}_\Lambda \left( \mathcal{F} C G_\theta(m_u|_\Theta) \right), d|_\Lambda \right) \right],
\end{equation}
where $\mathcal{P}_\Lambda$ is a projection onto the supervision set in k-space. This technique enables network training without ever accessing fully-sampled ground truth, making it particularly appealing in real-world clinical scenarios \cite{millard2023theoretical,desai2023noise2recon}.

\subsubsection{Scan-Specific Training}
When training data are entirely absent due to limited scanning resources, network models must be learned exclusively on test data during inference \cite{RAKI,kim2019loraki,arefeen2022mrm}. Note that such scan-specific learning can be categorized as a subclass of self-supervised techniques that learn from a single test sample (i.e., a single test subject). A prominent approach uses a randomly-initialized unconditional network $G_\theta$ to map a fixed noise input $z$ onto an image-domain reconstruction \cite{Jin2019,Arora2020ismrm,darestani2021accelerated,slavkova2023untrained}. The network is optimized to satisfy data consistency in k-space based on the available undersampled acquisition, as inspired by deep image prior (DIP) techniques \cite{ulyanov_deep_2020} (Fig. \ref{fig:dl_scheme}c.1):
\begin{equation}
\mathcal{L}_{\text{DIP}}(\theta) = \ell \left( \mathcal{P}_\Omega \left( \mathcal{F} C G_\theta(z) \right), d \right),
\end{equation}
where $\mathcal{P}_\Omega$ denotes the sampling mask. 

Another approach uses a randomly-initialized conditional network $G_\theta$ to map a linear transformation of a subset $\Theta$ of undersampled data (e.g., the zero-filled Fourier reconstruction $m_u|_\Theta$ of $d|_\Theta$ onto a reconstruction \cite{arefeen2022mrm,yaman2022zeroshot}. In this case, the network is optimized to satisfy consistency to another subset $\Lambda$ of undersampled data (Fig. \ref{fig:dl_scheme}c.2):\footnote{An implementation of a scan-specific model is available at \url{https://github.com/tutorial-MRI-recon/tutorial/tree/main/scan_specific_dl}.}
\begin{equation}
\mathcal{L}_{\text{ZS}}(\theta) = \ell \left( \mathcal{P}_\Lambda \left( \mathcal{F} C G_\theta(m_u |_\Theta) \right), d|_\Lambda \right).
\label{eq:ZStraining}
\end{equation}

Test-time trained networks can adapt to the unique characteristics of each scan, albeit running optimization during inference is computationally expensive. Moreover, when the entire set of acquired data is used during training, the network might suffer from overfitting to noise if not early stopped \cite{wang2023early}. A promising strategy to alleviate this issue is to hold out an independent subset of k-space samples as validation data to guide early stopping \cite{yaman2022zeroshot}. For instance, a third disjoint set of k-space data $\mathcal{V}$ (i.e., $\Theta \cup \Lambda \cup \mathcal{V}= \Omega$) can be reserved to assess the validation performance of the conditional network described in (\ref{eq:ZStraining}).

\subsubsection{Hybrid Training}
A network learned on a training dataset collected at a particular imaging site or with a specific protocol might later be deployed in different settings, yielding shifts in the data distribution. These shifts can in turn cause poor generalization in learned networks, necessitating domain alignment. Hybrid training combines the strengths of population-level and scan-specific paradigms to enhance generalization on out-of-distribution test data \cite{tezcan2018mr,Biswas2019,korkmaz2022unsupervised,alccalar2024zero}. In this framework, a network $G_{\theta_0}$ initially trained on a cohort of training subjects is later adapted on the undersampled test data from a specific subject (Fig. \ref{fig:dl_scheme}c.2):
\begin{equation}
\mathcal{L}{_\text{hybrid}}(\theta) = \ell \left( \mathcal{P}_{\Omega} \left( \mathcal{F} C G_\theta(z) \right), d \right) \quad \text{with} \quad \theta \leftarrow \theta_0,
\end{equation}
where $z$ is a noise variable or an aliased image, $\theta$ is initialized from $\theta_0$. Hybrid training leverages population priors while retaining the adaptability of scan-specific optimization.

\begin{figure*}[t]
  \centering
  \vspace{-0.1cm}
  \includegraphics[width=\textwidth]{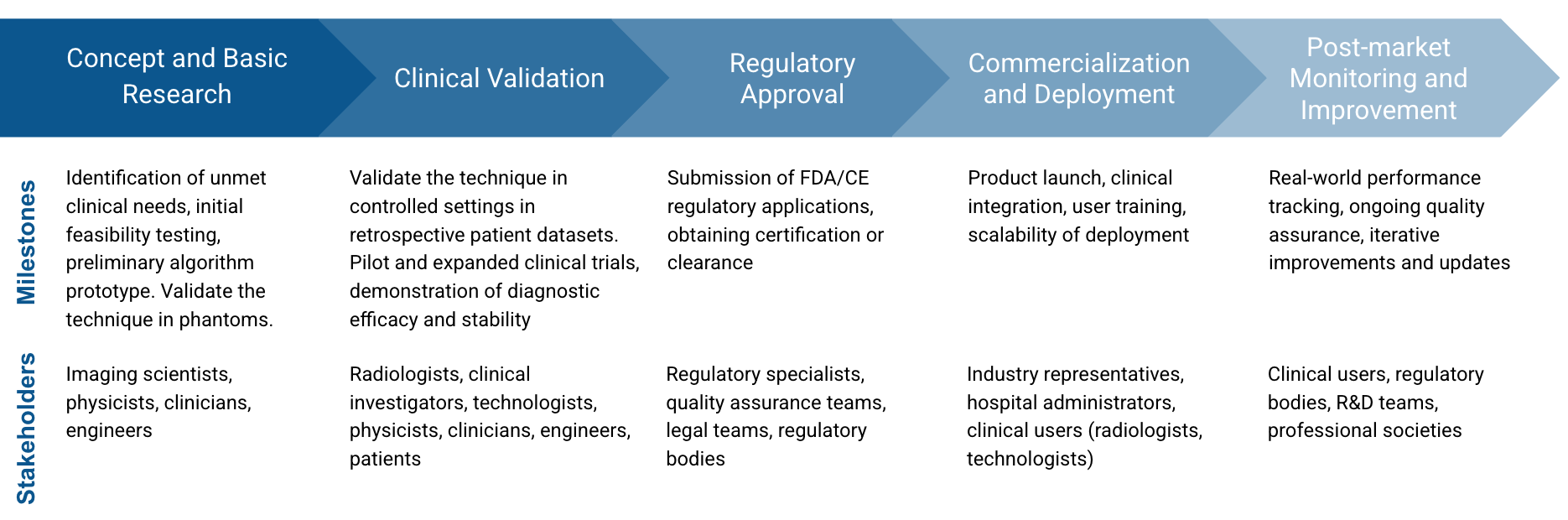}
  \caption{Stages of translational development for MRI reconstruction techniques, alongside the critical milestones and involved stakeholders.}
  \label{fig:translational_stages}
\end{figure*}

\section{Clinical Implications and Translation}

\subsection{General Implications}
\subsubsection{Improved Image Quality for Visual Assessment}
Technical advances in MRI encoding and reconstruction have enabled recovery of high-fidelity images from heavily undersampled data, which is particularly impactful in clinical settings where scan time is limited but the need for diagnostic accuracy is paramount \cite{Levin2016-sk}. In essence, improvements in image quality—such as enhanced contrast, reduced noise, and clearer delineation of anatomical structures—enable radiologists to make more accurate and confident interpretations, particularly in complex or borderline cases. For instance, high tissue contrast and structure preservation contribute to improved detection and delineation of pathological findings such as ischemic lesions, demyelinating plaques, or small tumors \cite{Kiryu2023-ju}. The extent of the lesion delineation is also important \cite{Peng2024-jt}. In addition, improved consistency in image texture and sharpness across slices and contrasts contributes to a more coherent visual impression \cite{Mooney2019-de}. This uniformity reduces cognitive load on the reader, minimizes interpretation time, and decreases intra-reader and inter-reader variability \cite{Peng2024-jt}.

\subsubsection{Enhanced Patient Compliance and Comfort}
Accelerated MRI reconstruction techniques offer benefits not only for image quality but also for patient compliance and comfort \cite{lustig2007sparse,yoo2023deep}. One of the most direct advantages is the reduction of scan duration. Shorter scans can minimize the time patients are required to remain still inside the scanner, thereby reducing physical discomfort, especially for populations such as children, elderly individuals, or critically ill patients \cite{Afacan2016-hi,Andre2015-fk}.

Increased efficiency in MRI scanning schedules improves throughput and access to imaging services. Shortened scan durations also decrease the likelihood of motion-related artifacts, which are one of the most common sources of non-diagnostic or inconclusive scans \cite{Zaitsev2015-sq}. In high-demand clinical settings, minimizing the need for repeat acquisitions and completing exam protocols within limited time windows help alleviate scheduling bottlenecks, reduce patient wait times, lower operational costs, and enhance the overall patient experience \cite{Nguyen2022-yc}.

These safety and comfort gains, while sometimes considered secondary to diagnostic efficacy, are in fact essential to patient-centered care. They contribute to higher compliance rates, fewer aborted scans, and an overall improvement in the quality of care delivered in the imaging suite.

\subsubsection{Expanded Access to Advanced Imaging Protocols}
The integration of advanced MRI reconstruction techniques, particularly those enabling high acceleration factors, has the potential to allow for the incorporation of additional or more complex pulse sequences that would otherwise be excluded due to time constraints. This is particularly significant for multi-parametric, multi-contrast, or dynamic imaging protocols that can provide more comprehensive diagnostic information. For instance, advanced reconstructions can facilitate the routine inclusion of diffusion, perfusion, spectroscopy, or relaxometry—all of which are often omitted in time-limited clinical workflows despite their diagnostic value \cite{Seiberlich2020-xs}. In neuroimaging, this may enable broader application of multi-shell diffusion protocols or resting-state functional MRI, while in body imaging, time-resolved sequences (e.g., GRASP) may become more feasible \cite{Feng2016-nu}.

From a practical perspective, this expansion can contribute to earlier detection of disease, improved lesion characterization, and more precise treatment planning. Furthermore, it can help bring research-grade imaging techniques into standard clinical practice, thereby narrowing the translational gap between academic innovation and real-world diagnostics \cite{Towfighi2020-wu}. Additionally, advanced protocols supported by fast acquisitions are particularly valuable in longitudinal studies or follow-up examinations, where consistent but fast imaging is critical to monitor disease progression or treatment response. Ultimately, the capacity to deliver richer imaging datasets within standard clinical time slots represents a major step toward personalized and precision medicine \cite{Unknown2024-eh}.

\subsubsection{Facilitation of Data-Demanding Analysis Pipelines}
As modern radiology increasingly relies on computational tools for image processing and decision support, the role of image reconstruction extends beyond generating visually interpretable images. It now serves as the foundational step in a data pipeline that feeds into downstream applications such as automated segmentation, radiomics, disease classification, and predictive modeling \cite{Rao2025-df,Hagiwara2023-kr}.

In addition, accelerated reconstruction techniques facilitate the acquisition of large volumes of imaging data without exceeding clinically acceptable scan durations. This is especially relevant for 3D or 4D protocols, multi-contrast studies, or high-resolution imaging where large data throughput is needed for training and inference with DL models. For example, in radiomic workflows, the variability in input quality due to undersampled acquisitions can be mitigated by advanced reconstruction algorithms, thereby enhancing the reproducibility and robustness of the extracted features \cite{van-Timmeren2020-ar,Hagiwara2020-lr}. In this context, the synergy between advanced reconstruction and data-centric analysis tools forms a critical axis for the future of precision radiology \cite{Paverd2024-tm}.

\subsection{Application-Specific Implications}
Although many current clinical MRI protocols employ modest acceleration supported by classical reconstruction methods, e.g. R$=$2-4 using PI, substantial opportunities remain to further improve scan efficiency via adoption of advanced reconstruction techniques. Benefits of additional acceleration would be particularly evident in longer protocols, where reduced scan times can significantly improve workflow, enhance patient comfort, and lower the risk of motion-related artifacts \cite{yvonnelui_review}. The clinical value of MRI reconstruction should therefore be evaluated in relation to exam durations and diagnostic requirements across anatomical regions and applications.

\textbf{Neuro MRI.} In neuroimaging, standard clinical protocols are designed to acquire multiple image contrasts, such as T\textsubscript{1}-weighted, T\textsubscript{2}-weighted, FLAIR, diffusion, and susceptibility-weighted imaging, typically taking 20 to 30 minutes \cite{knoll2020deep}. Advanced reconstruction techniques can enable higher acceleration rates to be used, reducing the scan time to as little as 5 to 10 minutes. Protocols that involve high-resolution volumetric imaging across multiple contrasts, such as those used in epilepsy evaluation (usually 30 to 40 minutes \cite{Conti2024-am}), stand to benefit significantly. In addition to improving patient compliance, especially among pediatric or neurologically impaired populations, the time saved can be reinvested to increase spatial resolution or to include additional contrasts, thereby enhancing lesion conspicuity and anatomical detail without prolonging the overall exam duration.

\textbf{MSK MRI.} In musculoskeletal imaging, fine anatomical delineation is required for assessing small structures such as ligaments, tendons, cartilage, or subchondral bone \cite{Flores2023-bi}. Due to the requirement for high spatial resolution and multiple oblique imaging planes aligned with complex joint anatomy, musculoskeletal imaging typically takes 15 to 30 minutes \cite{johnson2023deep}. The time saved with higher acceleration using advanced reconstruction techniques can be reinvested to improve SNR or further enhance spatial resolution, thereby facilitating detection of subtle findings such as partial tendon tears, cartilage defects, and marrow lesions.

\textbf{Abdominal MRI.} In abdominal imaging, challenges related to respiratory motion and contrast timing often necessitate extended protocols over multiple breath-holds, lasting 30–45 minutes \cite{feng2017compressed}. Free-breathing acquisitions combined with advanced reconstruction methods can reduce scan times to 10–15 minutes. This can enable more consistent image quality across patients, mitigate motion artifacts, and preserve key contrasts such as lesion enhancement. For prostate exams, high quality T\textsubscript{2}-weighted images in multiple planes and DWI are often required, while contrast enhanced scans are also used in subpopulations \cite{Purysko2020-mo}. Reconstruction techniques can enable shorter and/or higher-resolution scans, improving lesion detection while streamlining screening and diagnostic workflows.

\textbf{Cardiac MRI.} Used pervasively in evaluation of cardiovascular diseases, cardiac MRI involves detailed assessments across multiple domains, such as cardiac morphology and function, myocardial tissue characterization, perfusion imaging, flow and hemodynamics, or coronary artery evaluation \cite{Rajiah2023-mz}. These assessments are encoding-intensive due to the need for dynamic, multi-phase, and multi-contrast data acquisitions. Additionally, respiratory and cardiac motion often necessitate multiple breath-holds during scanning. The average scan duration ranges from 30 to 60 minutes, depending on protocol complexity (e.g., inclusion of perfusion, late gadolinium enhancement, or mapping sequences) and patient-related factors such as breath-holding capacity and the presence of arrhythmias \cite{oscanoa2023deep,morales2024present}.
Reconstruction techniques can facilitate the use of higher acceleration and reduce scan time to 10–20 minutes, enabling higher temporal resolution or more effective motion correction. These improvements can enhance the visualization of wall motion, myocardial strain, and perfusion dynamics.

\textbf{Whole-body MRI.} In whole-body imaging, increasingly adopted in oncology for metastatic screening and treatment monitoring, protocols often extend to 45–90 minutes \cite{lecouvet2024present}. Advanced acquisition/reconstruction, coupled with motion-compensation techniques, can reduce exam times to 15–30 minutes without compromising diagnostic accuracy.

\textbf{MR-guided Radiotherapy.} MR-guided radiotherapy leverages MRI’s superior soft-tissue contrast and real-time imaging for precise tumor targeting and on-table adaptive planning without ionizing radiation \cite{Pollard2017-wj, De-Pietro2024-mk}. Yet, cine MR sequences for gating and intrafraction tracking require high temporal resolution, minimal latency, and robustness to motion \cite{Otazo2021-no}. Advanced reconstruction methods can provide low-latency image formation with improved artifact suppression, supporting reliable margin definition in abdominal and thoracic sites and enhancing treatment precision.

\textbf{Interventional MRI.} Guiding procedures such as biopsies, ablations, and catheter-based therapies \cite{Thompson2021-sr}, interventional MRI requires rapid, high-temporal-resolution visualization of a limited field of view to track instruments and target tissues in real time. Frequent presence of needles, wires, and ablation devices further necessitates robust artifact mitigation \cite{Campbell-Washburn2019-ub}. Advanced reconstructions that accelerate imaging with minimal latency while preserving spatial fidelity can support continuous procedural monitoring, faster intra-procedural decision-making, and safer, more effective interventions overall.

\textbf{Low-field MRI.} Low-field MRI ($\lesssim$0.5T) offers reduced equipment cost, portability, and improved safety for patients with implants, pediatric populations, or those in resource-limited or bedside settings \cite{shimronlowfield}. However, lower SNR and reduced spatial resolution are common limitations at these field strengths that can compromise diagnostic utility. Advanced reconstruction methods can mitigate these challenges by improving scan efficiency, lowering image noise, and improving spatial resolution.

\textbf{Others.} Finally, emerging applications such as lung, breast, and fetal MRI stand to benefit substantially from advanced reconstruction. These techniques can help address longstanding barriers including low signal, motion artifacts, and time constraints. As such methods continue to mature, they may drive broader clinical adoption and enable more accessible, efficient MRI workflows for screening and point-of-care evaluations.

\begin{table*}[t]
\centering
\caption{Categorization of MRI reconstruction methods and considerations for their clinical translation.}
\label{tab:MRI_Reconstruction_Classification}
\begin{tabular}{p{1.25cm} p{1.25cm} p{3.7cm} p{3.7cm} p{3.7cm} p{2.0cm}}
\toprule
\textbf{Category} & \textbf{Methods} & \textbf{Key Characteristics} & \textbf{Pros} & \textbf{Cons} & \textbf{Current Translational Stage} \\ 
\midrule
\multirow[t]{4}{*}{%
    \parbox[t]{1.5cm}{%
        \raggedright
        \textbf{Classical}\\
        \textbf{Methods}
    }%
}  & \raggedright Linear Regression / Interpolation & Exploits coil-sensitivity encoding; straightforward implementation; clinically established. & Computationally efficient implementation; clinically accepted; minimal regulatory hurdles. & Limited acceleration (R), residual artifacts/noise towards higher R.& Post-market Monitoring and Improvement \\
\cmidrule{2-6}
& Sparsity & Exploits transform-domain sparsity; iterative optimization-based implementation; broad applicability across MRI sequences. &  Moderate R, strong artifact reduction, clinically validated. &  High computational cost due to iterative optimization; requires careful parameter tuning. & Commercialization and Deployment Stage \\
\cmidrule{2-6}
& Low-Rank &  Exploits structured low-rankness; especially beneficial on dynamic MRI data. & Decent artifact suppression; well-suited for data with significant spatial/temporal correlations. & Computationally intensive; relatively complex, non-standard implementation. & Mainly Concept and Basic Research \\
\cmidrule{2-6}
& Subspace & Utilizes Bloch-simulated signal models; beneficial in parameter mapping applications. & Quantitative imaging capability; high flexibility in post-hoc contrast generation. & Complexity of dictionary calibration; longer development cycles. & Some are FDA/CE approved \\
\midrule
\multirow[t]{4}{*}{%
    \parbox[t]{1.5cm}{%
        \raggedright
        \textbf{Deep-Learning\\
        Methods}
    }%
} 
& Data-Driven & Leverages priors learned from exemplar data; reconstruction cast as a single forward pass through the network. & High reconstruction quality; rapid inference; minimal parameter tuning. & Limited out-of-domain generalization; regulatory complexities due to black-box nature. & Commercialization and Deployment Stage \\
\cmidrule{2-6}
& Physics-Guided & Combines data- and physics-driven priors; reconstruction cast as iterations through the two types of priors. & Better physical grounding; higher user trust; improved generalization under changes to the imaging operator. & Higher computational burden than data-driven methods; limited structural details in reconstructed images. & Mainly Clinical Validation, some are FDA/CE approved \\
\cmidrule{2-6}
& Generative & Leverages task-agnostic generative priors to synthesize MR images; enhances structural details. &  High visual quality; excellent detail preservation. & Susceptibility to hallucinations; reduced user trust; typically heavy computational burden. & Mainly Concept and Basic Research \\

\bottomrule
\end{tabular}
\end{table*}

\subsection{Stages of Translational Development}
The translation of MRI reconstruction technologies into clinical practice generally follows a multi-stage pipeline. Although several definitions of translational research phases have been proposed in the biomedical sciences, especially in genomics, they may not fully capture the specific challenges associated with imaging technologies \cite{Gannon2014-az,fort2017mapping}. As summarized in Fig.~\ref{fig:translational_stages}, the key stages include conceptual development, clinical validation, regulatory approval, commercialization, and post-market monitoring.

In the early conceptual stage, reconstruction methods are developed and tested on simulated or phantom data to establish feasibility. Subsequent clinical validation involves retrospective evaluation on patient datasets and prospective pilot studies, aiming to demonstrate diagnostic efficacy, reproducibility, and robustness in real-world settings. Most studies leverage image fidelity metrics such as MSE, PSNR, and SSIM, which are widely used in computer vision, to provide a quantitative benchmark of reconstruction performance. While valuable for assessing basic image similarity, these conventional metrics can have limited correspondence to clinical utility and should be interpreted with caution when used as indicators of diagnostic value. Ideally, metrics that better align with human perception of image quality can be incorporated to improve fidelity of metric-based evaluations \cite{Reisenhofer2018-ni, Karner2025-gf}. More importantly, clinical validation should not rely solely on quality metrics but rather on demonstrated clinical value, that is, whether the reconstruction supports accurate diagnosis and appropriate patient management \cite{Knoll2020-de}.

A particularly challenging step in this pipeline is regulatory approval. MRI reconstruction techniques, especially those involving machine learning, may be subject to rigorous review by agencies such as the U.S. Food and Drug Administration (FDA) or European CE marking authorities. Demonstrating safety, interpretability, and generalizability is essential.

Following regulatory clearance, commercialization and deployment require integration into clinical workflows and imaging platforms. This involves not only technical compatibility with PACS (picture archiving and communication systems) and scanner software, but also attention to computational demands, user training, and support infrastructure as scalable deployment demands efficient algorithms.

Finally, post-market surveillance is critical to ensure continued safety and efficacy, identify performance issues across diverse populations or institutions, and guide iterative updates \cite{Center-for-Devices2024}. Collaborative efforts between industry, regulatory bodies, and clinical users are essential to support this continuous improvement process. Overall, the successful translation of advanced MRI reconstruction techniques depends on coordinated progress across technical innovation, regulatory navigation, and system-level implementation.

\subsection{Clinical Workflow Integration}
Despite their diagnostic potential, advanced reconstruction techniques face several challenges in clinical implementation. Each technique has its own strengths and limitations as briefly summarized in Table~\ref{tab:MRI_Reconstruction_Classification}, and it is important to understand these characteristics to select the most appropriate method for the task at hand. Beyond implications directly related to the level of acceleration, artifact suppression, and image quality, an important challenge in clinical implementation is the computational demand of advanced reconstruction methods. Integration of these methods into routine workflows inevitably depends on infrastructure readiness. High-performance computing resources, often including GPUs (graphical processing units), are typically required—posing constraints for on-scanner deployment. Off-scanner solutions, such as PACS-integrated servers or secure local clusters, provide a feasible alternative, but may introduce latency and complicate data routing.

Seamless adoption also demands interoperability with hospital IT (information technology) systems and compliance with DICOM (digital imaging and communications in medicine) standards. Deployment must be planned with attention to institutional regulations, particularly in regions with strict data governance policies. These operational factors underscore the importance of aligning algorithm development with practical implementation pathways. Effective integration requires not only technical compatibility, but also organizational adaptation and clinician engagement. Technologists and radiologists must be trained to understand both the capabilities and limitations of novel reconstruction methods. Trust in these systems is reinforced through transparent validation, consistent performance, and the ability to recognize artifacts or failure modes.

\section{Future Directions}
Learning-based MRI reconstruction has achieved impressive gains in scan acceleration and image quality in recent years. However, as modern methods transition from research to widespread clinical use, new challenges emerge that go beyond reconstruction performance. Future progress must also prioritize reliability and adaptability across diverse patient populations and clinical settings. Key areas of focus include mitigating the risk of hallucinations and bias, designing learning paradigms that scale with limited or heterogeneous data, and ensuring that methods are computationally efficient and infrastructure-compatible for real-world deployment.

\subsection{Enhancing Reliability Against Hallucinations}
Advances in MRI reconstruction have enabled the recovery of high-fidelity images from heavily undersampled acquisitions, significantly improving scan efficiency. To suppress aliasing artifacts in such acquisitions, reconstruction methods balance the support of acquired data with that of the imposed prior. As acceleration rates increase and fewer data are collected, reconstructions become more susceptible to hallucinations—manifesting as either loss or leakage of tissue features \cite{hallucination}. Loss occurs when true features cannot be recovered due to aliasing, while leakage refers to the appearance of features not present in the actual anatomy. Since hallucinatory features can closely resemble realistic anatomy, they pose serious risks for diagnostic reliability \cite{shimron2022implicit}.

Classical methods and some scan-specific DL methods rely on hand-crafted priors, which may lead to visible losses such as over-smoothing or signal attenuation in fine structures—potentially alleviated by tuning down the regularization strength at the expense of weaker artifact suppression. In contrast, non-scan-specific DL methods often learn from population data, risking leakage of features commonly seen in the pool of training subjects but absent in the test subject. A promising direction to mitigate hallucinations is the integration of priors grounded in physical signal models, resulting in hybrid approaches that combine classical and DL methods \cite{sriram2020grappanet,ISTA-NET,Primal_dual}. Ensemble approaches that aggregate predictions from diverse DL methods (e.g., physics-guided and generative networks) might also offer a level of safeguard by blending complementary strengths \cite{dar2021fewshot,ssdiffrecon}. Additionally, joint reconstruction strategies for multiple MRI contrasts can help enhance anatomical consistency across sequences, reducing the likelihood of hallucinated features appearing in any single contrast when proper measures are taken to avoid cross-contrast leakages \cite{Aggarwal2020-nd,polak2020joint}. Incorporating explicit anatomical constraints such as shape-aware priors might further regularize the solution space to favor plausible reconstructions \cite{WELLER2018118}.

While hallucinations may not be entirely avoidable, providing uncertainty or hallucination maps alongside reconstructed images \cite{edupuganti2020uncertainty,anastasiotmi2021} or using quantitative metrics that may be sensitive to presence of hallucinatory features \cite{hallumetric} can guide interpretation. Moreover, explainability tools—such as layer-wise relevance propagation \cite{bach2015pixel}, counterfactual generation \cite{bedel2023dreamr}, or network gradients \cite{sundararajan2017axiomatic}—can help identify which components of the network are responsible for hallucinations. Such retrospective analyses may support decisions about whether clinically relevant features are real or hallucinated. 

\subsection{Mitigating Biases from Data and Methodology}
As DL methods become increasingly prevalent in medical imaging, concerns about algorithmic bias have gained growing attention \cite{AIbias}. One major source of bias arises when reconstruction models, often trained on datasets that overrepresent certain populations or anatomical characteristics, systematically underperform on underrepresented groups \cite{du2023}. DL-based reconstruction models may inadvertently encode population-level priors that favor the majority distributions in the training data—at the expense of generalizability. Unlike hallucinations, which typically manifest as localized errors such as feature leakage or loss, bias often presents more subtly as systematic degradations in SNR, resolution, or contrast. It should be recognized that biases are not exclusive to DL methods—classical reconstructions can also produce characteristic artifacts due to over-emphasis of image features compatible with selected hand-crafted priors, such as block artifacts in total variation regularization. These biases can lead to inequities in image quality, diagnostic accuracy, and clinical decision-making.

Another important source of bias affecting the fidelity of model evaluation arises from shortcomings in methodological practices \cite{shimron2022implicit}. Common issues include assessing reconstruction models on MRI data that have been preprocessed in ways that simplify the underlying inverse problem—for example, using magnitude-only, coil-combined, or compressed images during modeling. Such simplifications can lead to overly optimistic performance metrics and poor generalization when models are eventually deployed on real-world MRI data, which are inherently complex and multi-coil.

Addressing bias in MRI reconstruction therefore requires a multifaceted strategy that combines data-centric and model-centric solutions. On the data side, curating diverse and representative datasets across demographics that contain raw, multi-coil, complex MRI data with minimal processing is fundamental. Collaborative learning frameworks offer a promising avenue to enhance training diversity while preserving patient privacy \cite{guo2021,FedGAT}. On the model side, advances in domain adaptation and model personalization can help mitigate performance degradations under domain shifts \cite{DINSDALE2021,dalmaz2024one}. Evaluating reconstruction performance across predefined subgroups—by demographic attributes or clinical conditions—is equally critical to reveal hidden disparities. Fairness-aware evaluation frameworks may provide systematic tools to quantify and address potential biases \cite{xu_addressing_2024}. Ultimately, transparent reporting of dataset composition, preprocessing steps, and model development/evaluation settings will be essential for equitable adoption of DL-based MRI reconstruction.

\vspace{0.2cm}
\subsection{Learning Under Data Constraints}
By learning task-specific data representations, DL architectures can surpass classical methods in capturing complex signal structure during MRI reconstruction. However, many supervised DL methods require large training datasets consisting of paired undersampled and fully-sampled acquisitions, and are typically trained for specific anatomies, contrasts, or undersampling patterns \cite{MoDl,Mardani2019b,hammernik2018learning,schlemper2017deep}. Unfortunately, even for a given anatomy, MRI scanners operate with a number of distinct protocols that vary widely in contrast, resolution, and other acquisition parameters. As a result, these models can suffer from degraded generalization performance when inevitably applied to new settings that differ from their training distribution \cite{liu2019santis,han2018deep}. Curating diverse and representative training datasets is particularly challenging in clinical environments, and even more so for multi-contrast protocols \cite{fastmri}. This data dependency raises a significant barrier for deployment of supervised DL in general-purpose MRI reconstruction.

To address these challenges, several alternative learning strategies can be considered. Common approaches include unsupervised paradigms where generative models learn the distribution of fully-sampled data without requiring paired examples \cite{oh2020unpaired,lei2020wasserstein,ozbey2022unsupervised}, and self-supervised methods that exploit inherent redundancies in the undersampled data to guide training \cite{yaman2020}. More broadly, federated learning provides a framework for decentralized training across institutions, enabling collaborative model development without the need to share any MRI data \cite{elmas2022federated,levac2023federated}. In scenarios where collaboration is impractical, synthetic datasets generated using physically-informed signal models can help alleviate data scarcity \cite{syntheticdata}. 

In resource-constrained settings, adaptation of foundation models on limited fine-tuning sets, or the development of scan-specific models without separate training data might also prove effective. Foundation models, pre-trained on large-scale, diverse datasets, can offer improved generalization through fine-tuning with smaller domain-specific datasets or few-shot learning techniques \cite{sun2025foundation,dong2025foundation}. In contrast, scan-specific approaches typically perform reconstruction modeling on a single undersampled scan using the acquired data and the imaging operator \cite{arefeen2022mrm}. While this approach enhances generalization to the individual scan's characteristics and avoids distribution shifts, it often requires longer reconstruction times compared to pre-trained models, which can be a trade-off in time-sensitive clinical workflows.

\subsection{Resource-Aware Computational Design}
A critical consideration for deploying MRI reconstruction methods in practice is their computational cost. MRI technicians often need to assess scan quality in real-time to determine whether a scan is diagnostically adequate or needs to be repeated due to motion artifacts, improper slice placement, or other issues \cite{murphy2012fast}. Providing timely feedback is especially important in busy clinical environments or for motion-sensitive populations such as pediatric or elderly patients. Furthermore, MRI protocols frequently involve multi-contrast imaging with high-channel receiver arrays, resulting in large data volumes that pose significant computational burden—particularly when real-time or near-real-time reconstruction is desired \cite{zhang2013coil}.

Several strategies can help mitigate these challenges. Coil compression techniques can reduce channel dimensionality, for instance by projecting 32-channel data to 16 virtual coils, thereby accelerating subsequent processing \cite{zhang2013coil}. Similarly, compression along contrast dimensions can reduce redundancy in multi-contrast acquisitions \cite{ma2013magnetic}. Parallelized implementations leveraging multiple GPUs or distributed compute nodes can further accelerate workflows \cite{schaetz2017accelerated}. Alternatively, sequential reconstruction pipelines may be employed, wherein an initial reference contrast is reconstructed with light undersampling and then used as a prior to guide joint reconstruction of the remaining contrasts \cite{gong2015promise}. While this approach may not fully exploit inter-contrast correlations, it offers a practical trade-off between computational efficiency and image quality.

\section{Conclusion}
We reviewed a broad range of MRI reconstruction methods that incorporate prior knowledge to improve quality of images recovered from accelerated acquisitions. Modern DL methods show strong potential to shorten scan times, expand access to advanced imaging protocols previously limited to research settings, and provide higher-quality inputs for downstream analysis. However, their clinical translation depends on rigorous validation, clear regulatory frameworks, scalable computational designs, and ongoing collaboration with technologists and radiologists to identify failure modes and sustain user trust. Ultimately, realizing the full potential of these methods in routine patient care will require interdisciplinary efforts among researchers, clinicians, vendors, and regulators.

\bibliographystyle{IEEEtran}
\bibliography{refs,refs_shohei}

\begin{thebibliography}{100}
\providecommand{\url}[1]{#1}
\csname url@samestyle\endcsname
\providecommand{\newblock}{\relax}
\providecommand{\bibinfo}[2]{#2}
\providecommand{\BIBentrySTDinterwordspacing}{\spaceskip=0pt\relax}
\providecommand{\BIBentryALTinterwordstretchfactor}{4}
\providecommand{\BIBentryALTinterwordspacing}{\spaceskip=\fontdimen2\font plus
\BIBentryALTinterwordstretchfactor\fontdimen3\font minus
  \fontdimen4\font\relax}
\providecommand{\BIBforeignlanguage}[2]{{%
\expandafter\ifx\csname l@#1\endcsname\relax
\typeout{** WARNING: IEEEtran.bst: No hyphenation pattern has been}%
\typeout{** loaded for the language `#1'. Using the pattern for}%
\typeout{** the default language instead.}%
\else
\language=\csname l@#1\endcsname
\fi
#2}}
\providecommand{\BIBdecl}{\relax}
\BIBdecl

\bibitem{Bernstein2004HandbookSequences}
M.~A. Bernstein, K.~F. King, and X.~J. Zhou,
  \emph{\BIBforeignlanguage{English}{{Handbook of MRI pulse sequences}}}.\hskip
  1em plus 0.5em minus 0.4em\relax Academic Press, 2004.

\bibitem{de1992mr}
B.~De~Coene \emph{et~al.}, ``{MR of the brain using fluid-attenuated inversion
  recovery (FLAIR) pulse sequences.}'' \emph{AJNR Am. J. Neuroradiol.},
  vol.~13, no.~6, pp. 1555--1564, 1992.

\bibitem{hajnal1992use}
J.~V. Hajnal \emph{et~al.}, ``{Use of fluid attenuated inversion recovery
  (FLAIR) pulse sequences in MRI of the brain},'' \emph{J. Comput. Assist.
  Tomogr.}, vol.~16, no.~6, pp. 841--844, 1992.

\bibitem{moseley1990early}
M.~E. Moseley \emph{et~al.}, ``{Early detection of regional cerebral ischemia
  in cats: comparison of diffusion- and T2-weighted MRI and spectroscopy},''
  \emph{Magn. Reson. Med.}, vol.~14, no.~2, pp. 330--346, 1990.

\bibitem{koh2007diffusion}
D.-M. Koh and D.~J. Collins, ``{Diffusion-weighted MRI in the body:
  applications and challenges in oncology},'' \emph{AJR Am. J. Roentgenol.},
  vol. 188, no.~6, pp. 1622--1635, 2007.

\bibitem{de2010quantitative}
L.~De~Rochefort \emph{et~al.}, ``{Quantitative susceptibility map
  reconstruction from MR phase data using Bayesian regularization: validation
  and application to brain imaging},'' \emph{Magn. Reson. Med.}, vol.~63,
  no.~1, pp. 194--206, 2010.

\bibitem{haacke2004susceptibility}
E.~M. Haacke \emph{et~al.}, ``{Susceptibility weighted imaging (SWI)},''
  \emph{Magn. Reson. Med.}, vol.~52, no.~3, pp. 612--618, 2004.

\bibitem{belliveau1991functional}
J.~W. Belliveau \emph{et~al.}, ``Functional mapping of the human visual cortex
  by magnetic resonance imaging,'' \emph{Science}, vol. 254, no. 5032, pp.
  716--719, 1991.

\bibitem{ogawa1990oxygenation}
S.~Ogawa \emph{et~al.}, ``Oxygenation-sensitive contrast in magnetic resonance
  image of rodent brain at high magnetic fields,'' \emph{Magn. Reson. Med.},
  vol.~14, no.~1, pp. 68--78, 1990.

\bibitem{setsompop2013pushing}
K.~Setsompop \emph{et~al.}, ``{Pushing the limits of in vivo diffusion MRI for
  the Human Connectome Project},'' \emph{Neuroimage}, vol.~80, pp. 220--233,
  2013.

\bibitem{detre1990mrm}
J.~A. Detre \emph{et~al.}, ``Perfusion imaging,'' \emph{Magn. Reson. Med.},
  vol.~23, no.~1, pp. 37--45, 1992.

\bibitem{lin2007mrm}
F.-H. Lin \emph{et~al.}, ``{Sensitivity-encoded (SENSE) proton echo-planar
  spectroscopic imaging (PEPSI) in the human brain},'' \emph{Magn. Reson.
  Med.}, vol.~57, no.~2, pp. 249--257, 2007.

\bibitem{fessler_review}
J.~A. Fessler, ``{Model-based image reconstruction for MRI},'' \emph{IEEE
  Signal Process. Mag.}, vol.~27, no.~4, pp. 81--89, 2010.

\bibitem{hammernik2022physicsdriven}
K.~Hammernik \emph{et~al.}, ``{Physics-driven deep learning for computational
  magnetic resonance imaging: combining physics and machine learning for
  improved medical imaging},'' \emph{IEEE Signal Process. Mag.}, vol.~40,
  no.~1, pp. 98--114, 2023.

\bibitem{Sodickson1997}
D.~K. Sodickson and W.~J. Manning, ``Simultaneous acquisition of spatial
  harmonics ({SMASH}): fast imaging with radiofrequency coil arrays,''
  \emph{Magn. Reson. Med.}, vol.~38, no.~4, pp. 591--603, 1997.

\bibitem{SENSE}
K.~P. Pruessmann \emph{et~al.}, ``{SENSE}: sensitivity encoding for {fast
  MRI},'' \emph{Magn. Reson. Med.}, vol.~42, no.~5, pp. 952--962, 1999.

\bibitem{griswold2002generalized}
M.~A. Griswold \emph{et~al.}, ``Generalized autocalibrating partially parallel
  acquisitions ({GRAPPA}),'' \emph{Magn. Reson. Med.}, vol.~47, no.~6, pp.
  1202--1210, 2002.

\bibitem{phasedarray}
P.~B. Roemer \emph{et~al.}, ``{The NMR phased array},'' \emph{Magn. Reson.
  Med.}, vol.~16, no.~2, pp. 192--225, 1990.

\bibitem{ravi_review}
S.~Ravishankar, J.~C. Ye, and J.~A. Fessler, ``{Image reconstruction: from
  sparsity to data-adaptive methods and machine learning},'' \emph{Proc. IEEE},
  vol. 108, no.~1, pp. 86--109, 2020.

\bibitem{uecker_review}
X.~Wang \emph{et~al.}, ``Physics-based reconstruction methods for magnetic
  resonance imaging,'' \emph{Philos. Trans. R. Soc. A, Math. Phys. Eng. Sci.},
  vol. 379, no. 2200, p. 20200196, 2021.

\bibitem{liang_review}
D.~Liang \emph{et~al.}, ``Deep magnetic resonance image reconstruction: Inverse
  problems meet neural networks,'' \emph{IEEE Signal Process. Mag.}, vol.~37,
  no.~1, pp. 141--151, 2020.

\bibitem{lustig2007sparse}
M.~Lustig, D.~Donoho, and J.~M. Pauly, ``Sparse {MRI}: the application of
  compressed sensing for rapid {MR} imaging,'' \emph{Magn. Reson. Med.},
  vol.~58, no.~6, pp. 1182--1195, 2007.

\bibitem{block2007undersampled}
K.~T. Block, M.~Uecker, and J.~Frahm, ``Undersampled radial {MRI} with multiple
  coils. {Iterative} image reconstruction using a total variation constraint,''
  \emph{Magn. Reson. Med.}, vol.~57, no.~6, pp. 1086--1098, 2007.

\bibitem{akccakaya2010compressed}
M.~Ak{\c{c}}akaya \emph{et~al.}, ``Compressed sensing with wavelet domain
  dependencies for coronary {MRI}: a retrospective study,'' \emph{IEEE Trans.
  Med. Imaging}, vol.~30, no.~5, pp. 1090--1099, 2011.

\bibitem{murphy2012fast}
M.~Murphy \emph{et~al.}, ``Fast $\ell_1$-{SPIRiT} compressed sensing parallel
  imaging {MRI}: scalable parallel implementation and clinically feasible
  runtime,'' \emph{IEEE Trans. Med. Imaging}, vol.~31, no.~6, pp. 1250--1262,
  2012.

\bibitem{liang2009accelerating}
D.~Liang \emph{et~al.}, ``Accelerating {SENSE} using compressed sensing,''
  \emph{Magn. Reson. Med.}, vol.~62, no.~6, pp. 1574--1584, 2009.

\bibitem{feng2017compressed}
L.~Feng \emph{et~al.}, ``{Compressed sensing for body MRI},'' \emph{J. Magn.
  Reson. Imaging}, vol.~45, no.~4, pp. 966--987, 2017.

\bibitem{shin2014calibrationless}
P.~J. Shin \emph{et~al.}, ``Calibrationless parallel imaging reconstruction
  based on structured low-rank matrix completion,'' \emph{Magn. Reson. Med.},
  vol.~72, no.~4, pp. 959--970, 2014.

\bibitem{mani2017multi}
M.~Mani \emph{et~al.}, ``{Multi-shot sensitivity-encoded diffusion data
  recovery using structured low-rank matrix completion (MUSSELS)},''
  \emph{Magn. Reson. Med.}, vol.~78, no.~2, pp. 494--507, 2017.

\bibitem{LORAKS}
J.~P. Haldar, ``{Low-rank modeling of local $k$-space neighborhoods (LORAKS)
  for constrained MRI},'' \emph{IEEE Trans. Med. Imaging}, vol.~33, no.~3, pp.
  668--681, 2014.

\bibitem{ALOHA}
D.~Lee \emph{et~al.}, ``{Acceleration of MR parameter mapping using
  annihilating filter-based low rank {Hankel} matrix (ALOHA)},'' \emph{Magn.
  Reson. Med.}, vol.~76, no.~6, pp. 1848--1864, 2016.

\bibitem{sandino2020SPM}
C.~M. Sandino \emph{et~al.}, ``Compressed sensing: From research to clinical
  practice with deep neural networks: Shortening scan times for magnetic
  resonance imaging,'' \emph{IEEE Signal Process. Mag.}, vol.~37, no.~1, pp.
  117--127, 2020.

\bibitem{chen2022review}
Y.~Chen \emph{et~al.}, ``{AI-based reconstruction for fast MRI—A systematic
  review and meta-analysis},'' \emph{Proc. IEEE}, vol. 110, no.~2, pp.
  224--245, 2022.

\bibitem{heckel2024deep}
R.~Heckel \emph{et~al.}, ``{Deep learning for accelerated and robust MRI
  reconstruction},'' \emph{MAGMA}, vol.~37, no.~3, pp. 335--368, 2024.

\bibitem{Wang2016}
S.~Wang \emph{et~al.}, ``{Accelerating magnetic resonance imaging via deep
  learning},'' in \emph{{IEEE ISBI}}, 2016, pp. 514--517.

\bibitem{ADMM-CSNET}
Y.~{Yang} \emph{et~al.}, ``{ADMM-CSNet: A} deep learning approach for image
  compressive sensing,'' \emph{IEEE Trans. Pattern Anal. Mach. Intell.},
  vol.~42, no.~3, pp. 521--538, 2020.

\bibitem{hammernik2018learning}
K.~Hammernik \emph{et~al.}, ``Learning a variational network for reconstruction
  of accelerated {MRI data},'' \emph{Magn. Reson. Med.}, vol.~79, no.~6, pp.
  3055--3071, 2018.

\bibitem{Mardani2019b}
M.~Mardani \emph{et~al.}, ``{Deep generative adversarial neural networks for
  compressive sensing MRI},'' \emph{IEEE Trans. Med. Imaging}, vol.~38, no.~1,
  pp. 167--179, 2019.

\bibitem{schlemper2017deep}
J.~Schlemper \emph{et~al.}, ``A deep cascade of convolutional neural networks
  for {MR} image reconstruction,'' in \emph{{IPMI}}.\hskip 1em plus 0.5em minus
  0.4em\relax Springer, 2017, pp. 647--658.

\bibitem{MoDl}
H.~K. {Aggarwal}, M.~P. {Mani}, and M.~{Jacob}, ``{MoDL: Model-Based} deep
  learning architecture for inverse problems,'' \emph{IEEE Trans. Med.
  Imaging}, vol.~38, no.~2, pp. 394--405, 2019.

\bibitem{quan2018compressed}
T.~M. Quan, T.~Nguyen-Duc, and W.-K. Jeong, ``{Compressed sensing MRI
  reconstruction using a generative adversarial network with a cyclic loss},''
  \emph{IEEE Trans. Med. Imaging}, vol.~37, no.~6, pp. 1488--1497, 2018.

\bibitem{Variational_end2end}
A.~Sriram \emph{et~al.}, ``End-to-end variational networks for accelerated
  {MRI} reconstruction,'' in \emph{MICCAI}, 2020, pp. 64--73.

\bibitem{Biswas2019}
S.~Biswas, H.~K. Aggarwal, and M.~Jacob, ``Dynamic {MRI} using model-based deep
  learning and {SToRM }priors: {MoDL-SToRM},'' \emph{Magn. Reson. Med.},
  vol.~82, no.~1, pp. 485--494, 2019.

\bibitem{Dar2017}
S.~U.~H. Dar \emph{et~al.}, ``A transfer-learning approach for accelerated
  {MRI} using deep neural networks,'' \emph{Magn. Reson. Med.}, vol.~84, no.~2,
  pp. 663--685, 2020.

\bibitem{zhu2018image}
B.~Zhu \emph{et~al.}, ``Image reconstruction by domain-transform manifold
  learning,'' \emph{Nature}, vol. 555, no. 7697, pp. 487--492, 2018.

\bibitem{nath2020datadriven}
R.~Nath \emph{et~al.}, ``Accelerated phase contrast magnetic resonance imaging
  via deep learning,'' in \emph{IEEE ISBI}, 2020, pp. 834--838.

\bibitem{jones2010diffusion}
D.~K. Jones, Ed., \emph{{Diffusion MRI: Theory, methods, and
  applications}}.\hskip 1em plus 0.5em minus 0.4em\relax Oxford Univ. Press,
  2010.

\bibitem{uecker2014espirit}
M.~Uecker \emph{et~al.}, ``{ESPIRiT—an eigenvalue approach to autocalibrating
  parallel MRI: where SENSE meets GRAPPA},'' \emph{Magn. Reson. Med.}, vol.~71,
  no.~3, pp. 990--1001, 2014.

\bibitem{PISCO}
R.~A. Lobos, C.-C. Chan, and J.~P. Haldar, ``New theory and faster computations
  for subspace-based sensitivity map estimation in multichannel {MRI},''
  \emph{IEEE Trans. Med. Imaging}, vol.~43, no.~1, pp. 286--296, 2024.

\bibitem{ying2007mrm}
L.~Ying and J.~Sheng, ``{Joint image reconstruction and sensitivity estimation
  in SENSE (JSENSE)},'' \emph{Magn. Reson. Med.}, vol.~57, no.~6, pp.
  1196--1202, 2007.

\bibitem{uecker2008mrm}
M.~Uecker \emph{et~al.}, ``{Image reconstruction by regularized nonlinear
  inversion—Joint estimation of coil sensitivities and image content},''
  \emph{Magn. Reson. Med.}, vol.~60, no.~3, pp. 674--682, 2008.

\bibitem{ying2009mrm}
B.~Liu \emph{et~al.}, ``{Regularized sensitivity encoding (SENSE)
  reconstruction using Bregman iterations},'' \emph{Magn. Reson. Med.},
  vol.~61, no.~1, pp. 145--152, 2009.

\bibitem{otazo2010combination}
R.~Otazo \emph{et~al.}, ``{Combination of compressed sensing and parallel
  imaging for highly accelerated first-pass cardiac perfusion MRI},''
  \emph{Magn. Reson. Med.}, vol.~64, no.~3, pp. 767--776, 2010.

\bibitem{beck2009fast}
A.~Beck and M.~Teboulle, ``A fast iterative shrinkage-thresholding algorithm
  for linear inverse problems,'' \emph{SIAM J. Imaging Sci.}, vol.~2, no.~1,
  pp. 183--202, 2009.

\bibitem{otazo2015low}
R.~Otazo, E.~Candes, and D.~K. Sodickson, ``{Low-rank plus sparse matrix
  decomposition for accelerated dynamic MRI with separation of background and
  dynamic components},'' \emph{Magn. Reson. Med.}, vol.~73, no.~3, pp.
  1125--1136, 2015.

\bibitem{haldar2015autocalibrated}
J.~P. Haldar, ``Autocalibrated {LORAKS} for fast constrained {MRI}
  reconstruction,'' in \emph{IEEE ISBI}.\hskip 1em plus 0.5em minus 0.4em\relax
  IEEE, 2015, pp. 910--913.

\bibitem{haldar2016p}
J.~P. Haldar and J.~Zhuo, ``{P-LORAKS}: low-rank modeling of local k-space
  neighborhoods with parallel imaging data,'' \emph{Magn. Reson. Med.},
  vol.~75, no.~4, pp. 1499--1514, 2016.

\bibitem{kim2017loraks}
T.~H. Kim, K.~Setsompop, and J.~P. Haldar, ``{LORAKS makes better SENSE:
  phase-constrained partial Fourier SENSE reconstruction without phase
  calibration},'' \emph{Magn. Reson. Med.}, vol.~77, no.~3, pp. 1021--1035,
  2017.

\bibitem{bilgic2018improving}
B.~Bilgic \emph{et~al.}, ``Improving parallel imaging by jointly reconstructing
  multi-contrast data,'' \emph{Magn. Reson. Med.}, vol.~80, no.~2, pp.
  619--632, 2018.

\bibitem{ongie2017fast}
G.~Ongie and M.~Jacob, ``A fast algorithm for convolutional structured low-rank
  matrix recovery,'' \emph{IEEE Trans. Comput. Imaging}, vol.~3, no.~4, pp.
  535--550, 2017.

\bibitem{Ravishankar2011a}
S.~Ravishankar and Y.~Bresler, ``{MR image reconstruction from highly
  undersampled k-space data by dictionary learning},'' \emph{IEEE Trans. Med.
  Imaging}, vol.~30, no.~5, pp. 1028--1041, 2011.

\bibitem{tamir2017t2}
J.~I. Tamir \emph{et~al.}, ``{T2 shuffling: sharp, multicontrast, volumetric
  fast spin-echo imaging},'' \emph{Magn. Reson. Med.}, vol.~77, no.~1, pp.
  180--195, 2017.

\bibitem{ma2013magnetic}
D.~Ma \emph{et~al.}, ``Magnetic resonance fingerprinting,'' \emph{Nature}, vol.
  495, no. 7440, pp. 187--192, 2013.

\bibitem{zhao2018improved}
B.~Zhao \emph{et~al.}, ``{Improved magnetic resonance fingerprinting
  reconstruction with low-rank and subspace modeling},'' \emph{Magn. Reson.
  Med.}, vol.~79, no.~2, pp. 933--942, 2018.

\bibitem{cao2022optimized}
X.~Cao \emph{et~al.}, ``{Optimized multi-axis spiral projection MR
  fingerprinting with subspace reconstruction for rapid whole-brain
  high-isotropic-resolution quantitative imaging},'' \emph{Magn. Reson. Med.},
  vol.~88, no.~1, pp. 133--150, 2022.

\bibitem{sumpf2011jmri}
T.~J. Sumpf \emph{et~al.}, ``{Model-based nonlinear inverse reconstruction for
  T2 mapping using highly undersampled spin-echo MRI},'' \emph{J Magn. Reson.
  Imaging}, vol.~34, no.~2, pp. 420--428, 2011.

\bibitem{Yu2018c}
G.~Yang \emph{et~al.}, ``{DAGAN: Deep de-aliasing generative adversarial
  networks for fast compressed sensing MRI reconstruction},'' \emph{IEEE Trans.
  Med. Imaging}, vol.~37, no.~6, pp. 1310--1321, 2018.

\bibitem{wang2022b}
G.~Wang \emph{et~al.}, ``{B-spline parameterized joint optimization of
  reconstruction and k-space trajectories (BJORK) for accelerated 2D MRI},''
  \emph{IEEE Trans. Med. Imaging}, vol.~41, no.~9, pp. 2318--2330, 2022.

\bibitem{goodfellow2016deep}
I.~Goodfellow, Y.~Bengio, and A.~Courville, \emph{Deep learning}.\hskip 1em
  plus 0.5em minus 0.4em\relax MIT Press, 2016.

\bibitem{radhakrishna2023jointly}
C.~G. Radhakrishna and P.~Ciuciu, ``{Jointly learning non-Cartesian k-space
  trajectories and reconstruction networks for 2D and 3D MR imaging through
  projection},'' \emph{Bioengineering}, vol.~10, no.~2, p. 158, 2023.

\bibitem{ahmad2020plug}
R.~Ahmad \emph{et~al.}, ``Plug-and-play methods for magnetic resonance imaging:
  Using denoisers for image recovery,'' \emph{IEEE Signal Process. Mag.},
  vol.~37, no.~1, pp. 105--116, 2020.

\bibitem{liu2020rare}
J.~Liu \emph{et~al.}, ``{RARE: Image reconstruction using deep priors learned
  without groundtruth},'' \emph{IEEE J. Sel. Top. Signal Process.}, vol.~14,
  no.~6, pp. 1088--1099, 2020.

\bibitem{Zhou_2020_CVPR}
B.~Zhou and S.~K. Zhou, ``{DuDoRNet: Learning a dual-domain recurrent network
  for fast MRI reconstruction with deep T1 prior},'' in \emph{IEEE CVPR}, 2020.

\bibitem{polak2020joint}
D.~Polak \emph{et~al.}, ``{Joint multi-contrast variational network
  reconstruction (jVN) with application to rapid 2D and 3D imaging},''
  \emph{Magn. Reson. Med.}, vol.~84, no.~3, pp. 1456--1469, 2020.

\bibitem{sun2019deep}
L.~Sun \emph{et~al.}, ``{A deep information sharing network for multi-contrast
  compressed sensing MRI reconstruction},'' \emph{IEEE Trans. Image Process.},
  vol.~28, no.~12, pp. 6141--6153, 2019.

\bibitem{yaman2020}
B.~Yaman \emph{et~al.}, ``Self-supervised learning of physics-guided
  reconstruction neural networks without fully sampled reference data,''
  \emph{Magn. Reson. Med.}, vol.~84, no.~6, pp. 3172--3191, 2020.

\bibitem{Ryu2021-oy}
K.~Ryu \emph{et~al.}, ``Accelerated multicontrast reconstruction for synthetic
  {MRI} using joint parallel imaging and variable splitting networks,''
  \emph{Med. Phys.}, vol.~48, no.~6, pp. 2939--2950, 2021.

\bibitem{liu2021regularization}
X.~Liu \emph{et~al.}, ``{On the regularization of feature fusion and mapping
  for fast MR multi-contrast imaging via iterative networks},'' \emph{Magn.
  Reson. Imaging}, vol.~77, pp. 159--168, 2021.

\bibitem{ramzi2022nc}
Z.~Ramzi \emph{et~al.}, ``{NC-PDNet: A density-compensated unrolled network for
  2D and 3D non-Cartesian MRI reconstruction},'' \emph{IEEE Trans. Med.
  Imaging}, vol.~41, no.~7, pp. 1625--1638, 2022.

\bibitem{Lam2023}
F.~Lam, X.~Peng, and Z.-P. Liang, ``High-dimensional {MR} spatiospectral
  imaging by integrating physics-based modeling and data-driven machine
  learning: Current progress and future directions,'' \emph{IEEE Signal
  Process. Mag.}, vol.~40, no.~2, pp. 101--115, 2023.

\bibitem{combettes2011proximal}
P.~L. Combettes and J.-C. Pesquet, ``Proximal splitting methods in signal
  processing,'' in \emph{Fixed-point algorithms for inverse problems in science
  and engineering}.\hskip 1em plus 0.5em minus 0.4em\relax Springer, 2011, pp.
  185--212.

\bibitem{chan2016plug}
S.~H. Chan, X.~Wang, and O.~A. Elgendy, ``Plug-and-play {ADMM} for image
  restoration: Fixed-point convergence and applications,'' \emph{IEEE Trans.
  Comput. Imaging}, vol.~3, no.~1, pp. 84--98, 2016.

\bibitem{romano2017little}
Y.~Romano, M.~Elad, and P.~Milanfar, ``{The little engine that could:
  Regularization by denoising (RED)},'' \emph{SIAM J. Imaging Sci.}, vol.~10,
  no.~4, pp. 1804--1844, 2017.

\bibitem{ryu2019plug}
E.~Ryu \emph{et~al.}, ``Plug-and-play methods provably converge with properly
  trained denoisers,'' in \emph{ICML}.\hskip 1em plus 0.5em minus 0.4em\relax
  PMLR, 2019, pp. 5546--5557.

\bibitem{liu2019mantis}
F.~Liu, L.~Feng, and R.~Kijowski, ``{MANTIS: model-augmented neural network
  with incoherent k-space sampling for efficient MR parameter mapping},''
  \emph{Magn. Reson. Med.}, vol.~82, no.~1, pp. 174--188, 2019.

\bibitem{Mani2020Model-BasedMRI}
M.~P. Mani \emph{et~al.}, ``{Model-based deep learning for reconstruction of
  joint k-q under-sampled high resolution diffusion MRI},'' in \emph{IEEE
  ISBI}.\hskip 1em plus 0.5em minus 0.4em\relax IEEE, 2020, pp. 913--916.

\bibitem{tezcan2018mr}
K.~C. Tezcan \emph{et~al.}, ``{MR} image reconstruction using deep density
  priors,'' \emph{IEEE Trans. Med. Imaging}, vol.~38, no.~7, pp. 1633--1642,
  2019.

\bibitem{raut2024generativeaivisionsurvey}
G.~Raut and A.~Singh, ``{Generative AI in vision: A survey on models, metrics
  and applications},'' \emph{arXiv:2402.16369}, 2024.

\bibitem{narnhofer2019inverse}
D.~Narnhofer \emph{et~al.}, ``{Inverse GANs for accelerated MRI
  reconstruction},'' in \emph{Proc. SPIE}, vol. 11138.\hskip 1em plus 0.5em
  minus 0.4em\relax SPIE, 2019, pp. 381--392.

\bibitem{korkmaz2022unsupervised}
Y.~Korkmaz \emph{et~al.}, ``{Unsupervised MRI reconstruction via zero-shot
  learned adversarial transformers},'' \emph{IEEE Trans. Med. Imaging},
  vol.~41, no.~7, pp. 1747--1763, 2022.

\bibitem{jalaln2021nips}
A.~Jalal \emph{et~al.}, ``Robust compressed sensing {MRI} with deep generative
  priors,'' in \emph{NeurIPS}, vol.~34, 2021, pp. 14\,938--14\,954.

\bibitem{chung2022media}
H.~Chung and J.~C. Ye, ``Score-based diffusion models for accelerated {MRI},''
  \emph{Med. Image Anal.}, vol.~80, p. 102479, 2022.

\bibitem{uecker2023mrm}
G.~Luo \emph{et~al.}, ``{Bayesian MRI reconstruction with joint uncertainty
  estimation using diffusion models},'' \emph{Magn. Reson. Med.}, vol.~90,
  no.~1, pp. 295--311, 2023.

\bibitem{ssdiffrecon}
Y.~Korkmaz, T.~Cukur, and V.~M. Patel, ``{Self-supervised MRI reconstruction
  with unrolled diffusion models},'' in \emph{MICCAI}.\hskip 1em plus 0.5em
  minus 0.4em\relax Springer, 2023, pp. 491--501.

\bibitem{gungor2022adaptive}
A.~G{\"u}ng{\"o}r \emph{et~al.}, ``{Adaptive diffusion priors for accelerated
  MRI reconstruction},'' \emph{Med. Image Anal.}, vol.~88, p. 102872, 2023.

\bibitem{liang2025analysis}
S.~Liang \emph{et~al.}, ``{Analysis of deep image prior and exploiting
  self-guidance for image reconstruction},'' \emph{IEEE Trans. Comput.
  Imaging}, 2025.

\bibitem{Kwon2017}
K.~Kwon, D.~Kim, and H.~Park, ``{A parallel MR imaging method using multilayer
  perceptron},'' \emph{Med. Phys.}, vol.~44, no.~12, pp. 6209--6224, 2017.

\bibitem{RAKI}
M.~Ak{\c{c}}akaya \emph{et~al.}, ``{Scan-specific robust
  artificial-neural-networks for k-space interpolation (RAKI) reconstruction:
  database-free deep learning for fast imaging},'' \emph{Magn. Reson. Med.},
  vol.~81, no.~1, pp. 439--453, 2019.

\bibitem{dosovitskiyViT}
A.~Dosovitskiy \emph{et~al.}, ``An image is worth 16x16 words: Transformers for
  image recognition at scale,'' in \emph{ICLR}, 2021.

\bibitem{guo2022reconformer}
P.~Guo \emph{et~al.}, ``Reconformer: Accelerated {MRI} reconstruction using
  recurrent transformer,'' \emph{IEEE Trans. Med. Imaging}, 2024.

\bibitem{huang2022swin}
J.~Huang \emph{et~al.}, ``{Swin transformer for fast MRI},''
  \emph{Neurocomputing}, vol. 493, pp. 281--304, 2022.

\bibitem{feng2021task}
C.-M. Feng \emph{et~al.}, ``{Task transformer network for joint MRI
  reconstruction and super-resolution},'' in \emph{MICCAI}.\hskip 1em plus
  0.5em minus 0.4em\relax Springer, 2021, pp. 307--317.

\bibitem{qin2018convolutional}
C.~Qin \emph{et~al.}, ``{Convolutional recurrent neural networks for dynamic MR
  image reconstruction},'' \emph{IEEE Trans. Med. Imaging}, vol.~38, no.~1, pp.
  280--290, 2018.

\bibitem{Hosseini2020b}
S.~A.~H. {Hosseini} \emph{et~al.}, ``Dense recurrent neural networks for
  accelerated {MRI: History}-cognizant unrolling of optimization algorithms,''
  \emph{IEEE J. Sel. Top. Signal Process.}, vol.~14, no.~6, pp. 1280--1291,
  2020.

\bibitem{guo2021over}
P.~Guo \emph{et~al.}, ``{Over-and-under complete convolutional RNN for MRI
  reconstruction},'' in \emph{MICCAI}.\hskip 1em plus 0.5em minus 0.4em\relax
  Springer, 2021, pp. 13--23.

\bibitem{gu2023mamba}
A.~Gu and T.~Dao, ``Mamba: Linear-time sequence modeling with selective state
  spaces,'' \emph{arXiv:2312.00752}, 2023.

\bibitem{huang2025enhancing}
J.~Huang \emph{et~al.}, ``{Enhancing global sensitivity and uncertainty
  quantification in medical image reconstruction with Monte Carlo
  arbitrary-masked Mamba},'' \emph{Med. Image Anal.}, vol.~99, p. 103334, 2025.

\bibitem{Mambarecon}
Y.~Korkmaz and V.~M. Patel, ``{MambaRecon: MRI reconstruction with structured
  state space models},'' in \emph{IEEE/CVF WACV}, 2025, pp. 4142--4152.

\bibitem{Mambaroll}
B.~Kabas \emph{et~al.}, ``Physics-driven autoregressive state space models for
  medical image reconstruction,'' \emph{arXiv:2412.09331}, 2024.

\bibitem{I2IMamba}
O.~F. {Atli} \emph{et~al.}, ``{I2I-Mamba: Multi-modal medical image synthesis
  via selective state space modeling},'' \emph{arXiv:2405.14022}, 2024.

\bibitem{kspaceDLMRI}
Y.~Han, L.~Sunwoo, and J.~C. Ye, ``${k}$-space deep learning for accelerated
  {MRI},'' \emph{IEEE Trans. Med. Imaging}, vol.~39, no.~2, pp. 377--386, 2020.

\bibitem{fdb}
M.~U. Mirza \emph{et~al.}, ``{Learning Fourier-constrained diffusion bridges
  for {MRI} reconstruction},'' \emph{arXiv:2308.01096}, 2023.

\bibitem{KikiNet}
T.~Eo \emph{et~al.}, ``{KIKI-net:} cross-domain convolutional neural networks
  for reconstructing undersampled magnetic resonance images,'' \emph{Magn.
  Reson. Med.}, vol.~80, no.~5, pp. 2188--2201, 2018.

\bibitem{Wang2019}
S.~Wang \emph{et~al.}, ``{DIMENSION: Dynamic MR imaging with both k-space and
  spatial prior knowledge obtained via multi-supervised network training},''
  \emph{NMR Biomed.}, p. e4131, 2022.

\bibitem{Liu2020HighlyPriors}
Q.~Liu \emph{et~al.}, ``{Highly undersampled magnetic resonance imaging
  reconstruction using autoencoding priors},'' \emph{Magn. Reson. Med.},
  vol.~83, no.~1, pp. 322--336, 1 2020.

\bibitem{FedGAT}
V.~A. Nezhad \emph{et~al.}, ``Generative autoregressive transformers for
  model-agnostic federated {MRI} reconstruction,'' \emph{arXiv:2502.04521},
  2025.

\bibitem{wang2024review}
S.~Wang \emph{et~al.}, ``{Knowledge-driven deep learning for fast MR imaging:
  Undersampled MR image reconstruction from supervised to un-supervised
  learning},'' \emph{Magn. Reson. Med.}, vol.~92, no.~2, pp. 496--518, 2024.

\bibitem{gong2015promise}
E.~Gong \emph{et~al.}, ``{PROMISE: parallel-imaging and compressed-sensing
  reconstruction of multicontrast imaging using SharablE information},''
  \emph{Magn. Reson. Med.}, vol.~73, no.~2, pp. 523--535, 2015.

\bibitem{bilgic2011multi}
B.~Bilgic, V.~K. Goyal, and E.~Adalsteinsson, ``{Multi-contrast reconstruction
  with Bayesian compressed sensing},'' \emph{Magn. Reson. Med.}, vol.~66,
  no.~6, pp. 1601--1615, 2011.

\bibitem{xiang2019rec}
L.~Xiang \emph{et~al.}, ``Deep-learning-based multi-modal fusion for fast {MR}
  reconstruction,'' \emph{IEEE Trans. Biomed. Eng.}, vol.~66, no.~7, pp.
  2105--2114, 2019.

\bibitem{Do2020-xp}
W.-J. Do \emph{et~al.}, ``{Reconstruction of multicontrast MR images through
  deep learning},'' \emph{Med. Phys.}, vol.~47, no.~3, pp. 983--997, 2020.

\bibitem{Kopanoglu2020-iq}
E.~Kopanoglu \emph{et~al.}, ``Simultaneous use of individual and joint
  regularization terms in compressive sensing: Joint reconstruction of
  multi-channel multi-contrast {MRI} acquisitions,'' \emph{NMR Biomed.},
  vol.~33, no.~4, p. e4247, 2020.

\bibitem{dar2020prior}
S.~U. Dar \emph{et~al.}, ``{Prior-guided image reconstruction for accelerated
  multi-contrast MRI via generative adversarial networks},'' \emph{IEEE J. Sel.
  Top. Signal Process.}, vol.~14, no.~6, pp. 1072--1087, 2020.

\bibitem{Kim2018-qd}
K.~H. Kim, W.~Do, and S.~Park, ``Improving resolution of {MR} images with an
  adversarial network incorporating images with different contrast,''
  \emph{Med. Phys.}, vol.~45, no.~7, pp. 3120--3131, 2018.

\bibitem{liu2020multi}
X.~Liu \emph{et~al.}, ``{Multi-contrast MR reconstruction with enhanced
  denoising autoencoder prior learning},'' in \emph{ISBI}.\hskip 1em plus 0.5em
  minus 0.4em\relax IEEE, 2020, pp. 1--5.

\bibitem{rizzuti2022joint}
G.~Rizzuti, A.~Sbrizzi, and T.~Van~Leeuwen, ``{Joint retrospective motion
  correction and reconstruction for brain MRI with a reference contrast},''
  \emph{IEEE Trans. Comput. Imaging}, vol.~8, pp. 490--504, 2022.

\bibitem{oved2025deeplearningpersonalizedpriors}
T.~Oved \emph{et~al.}, ``{Deep learning of personalized priors from past MRI
  scans enables fast, quality-enhanced point-of-care MRI with low-cost
  systems},'' \emph{arXiv:2505.02470}, 2025.

\bibitem{knoll2020deep}
F.~Knoll \emph{et~al.}, ``{Deep-learning methods for parallel magnetic
  resonance imaging reconstruction: A survey of the current approaches, trends,
  and issues},'' \emph{IEEE Signal Process. Mag.}, vol.~37, no.~1, pp.
  128--140, 2020.

\bibitem{Tamir2019}
J.~I. Tamir, S.~X. Yu, and M.~Lustig, ``Unsupervised deep basis pursuit:
  {Learning} reconstruction without ground-truth data,'' in \emph{{Proc.
  ISMRM}}, 2019, p. 0660.

\bibitem{blumenthal2024mrm}
M.~Blumenthal \emph{et~al.}, ``{Self-supervised learning for improved
  calibrationless radial MRI with NLINV-Net},'' \emph{Magn. Reson. Med.},
  vol.~92, no.~6, pp. 2447--2463, 2024.

\bibitem{cole2021fast}
E.~K. Cole \emph{et~al.}, ``{Fast unsupervised MRI reconstruction without
  fully-sampled ground truth data using generative adversarial networks},'' in
  \emph{ICCVW}, 2021, pp. 3988--3997.

\bibitem{millard2023theoretical}
C.~Millard and M.~Chiew, ``{A theoretical framework for self-supervised MR
  image reconstruction using sub-sampling via variable density
  Noisier2Noise},'' \emph{IEEE Trans. Comput. Imaging}, vol.~9, pp. 707--720,
  2023.

\bibitem{desai2023noise2recon}
A.~D. Desai \emph{et~al.}, ``{Noise2Recon: Enabling SNR-robust MRI
  reconstruction with semi-supervised and self-supervised learning},''
  \emph{Magn. Reson. Med.}, vol.~90, no.~5, pp. 2052--2070, 2023.

\bibitem{kim2019loraki}
T.~H. Kim, P.~Garg, and J.~P. Haldar, ``{LORAKI: Autocalibrated recurrent
  neural networks for autoregressive MRI reconstruction in k-space},''
  \emph{arXiv:1904.09390}, 2019.

\bibitem{arefeen2022mrm}
Y.~Arefeen \emph{et~al.}, ``{Scan-specific artifact reduction in k-space
  (SPARK) neural networks synergize with physics-based reconstruction to
  accelerate MRI},'' \emph{Magn. Reson. Med.}, vol.~87, no.~2, pp. 764--780,
  2022.

\bibitem{Jin2019}
K.~H. Jin \emph{et~al.}, ``Time-dependent deep image prior for dynamic {MRI},''
  \emph{arXiv:1910.01684}, 2019.

\bibitem{Arora2020ismrm}
S.~Arora, V.~Roeloffs, and M.~Lustig, ``Untrained modified deep decoder for
  joint denoising and parallel imaging reconstruction,'' in \emph{Proc. ISMRM},
  2020, p. 3585.

\bibitem{darestani2021accelerated}
M.~Z. Darestani and R.~Heckel, ``{Accelerated MRI with un-trained neural
  networks},'' \emph{IEEE Trans. Comput. Imaging}, vol.~7, pp. 724--733, 2021.

\bibitem{slavkova2023untrained}
K.~P. Slavkova \emph{et~al.}, ``{An untrained deep learning method for
  reconstructing dynamic MR images from accelerated model-based data},''
  \emph{Magn. Reson. Med.}, vol.~89, no.~4, pp. 1617--1633, 2023.

\bibitem{ulyanov_deep_2020}
D.~Ulyanov, A.~Vedaldi, and V.~Lempitsky, ``Deep {Image} {Prior},'' \emph{Int.
  J. Comput. Vis.}, vol. 128, no.~7, pp. 1867--1888, 2020.

\bibitem{yaman2022zeroshot}
B.~Yaman, S.~A.~H. Hosseini, and M.~Akcakaya, ``Zero-shot self-supervised
  learning for {MRI} reconstruction,'' in \emph{ICLR}, 2022.

\bibitem{wang2023early}
H.~Wang \emph{et~al.}, ``{Early stopping for Deep Image Prior},'' \emph{Trans.
  Mach. Learn. Res.}, 2021.

\bibitem{alccalar2024zero}
Y.~U. Al{\c{c}}alar and M.~Ak{\c{c}}akaya, ``Zero-shot adaptation for
  approximate posterior sampling of diffusion models in inverse problems,'' in
  \emph{ECCV}.\hskip 1em plus 0.5em minus 0.4em\relax Springer, 2024, pp.
  444--460.

\bibitem{Levin2016-sk}
D.~C. Levin and V.~M. Rao, ``\BIBforeignlanguage{en}{Factors that will
  determine future utilization trends in diagnostic imaging},''
  \emph{\BIBforeignlanguage{en}{J. Am. Coll. Radiol.}}, vol.~13, no.~8, pp.
  904--908, 2016.

\bibitem{Kiryu2023-ju}
S.~Kiryu \emph{et~al.}, ``\BIBforeignlanguage{en}{Clinical impact of deep
  learning reconstruction in {MRI}},''
  \emph{\BIBforeignlanguage{en}{Radiographics}}, vol.~43, no.~6, 2023.

\bibitem{Peng2024-jt}
W.~Peng \emph{et~al.}, ``\BIBforeignlanguage{en}{Prospective and multi-reader
  evaluation of deep learning reconstruction-based accelerated rectal {MRI}:
  image quality, diagnostic performance, and reading time},''
  \emph{\BIBforeignlanguage{en}{Eur. Radiol.}}, vol.~34, no.~11, pp.
  7438--7449, 2024.

\bibitem{Mooney2019-de}
S.~W. Mooney, P.~J. Marlow, and B.~L. Anderson, ``\BIBforeignlanguage{en}{The
  perception and misperception of optical defocus, shading, and shape},''
  \emph{\BIBforeignlanguage{en}{Elife}}, vol.~8, Jul. 2019.

\bibitem{yoo2023deep}
H.~Yoo \emph{et~al.}, ``{Deep learning--based reconstruction for acceleration
  of lumbar spine MRI: a prospective comparison with standard MRI},''
  \emph{Eur. Radiol.}, vol.~33, no.~12, pp. 8656--8668, 2023.

\bibitem{Afacan2016-hi}
O.~Afacan \emph{et~al.}, ``\BIBforeignlanguage{en}{Evaluation of motion and its
  effect on brain magnetic resonance image quality in children},''
  \emph{\BIBforeignlanguage{en}{Pediatr. Radiol.}}, vol.~46, no.~12, pp.
  1728--1735, Nov. 2016.

\bibitem{Andre2015-fk}
J.~B. Andre \emph{et~al.}, ``\BIBforeignlanguage{en}{Toward quantifying the
  prevalence, severity, and cost associated with patient motion during clinical
  {MR} examinations},'' \emph{\BIBforeignlanguage{en}{J. Am. Coll. Radiol.}},
  vol.~12, no.~7, pp. 689--695, Jul. 2015.

\bibitem{Zaitsev2015-sq}
M.~Zaitsev, J.~Maclaren, and M.~Herbst, ``\BIBforeignlanguage{en}{Motion
  artifacts in {MRI}: A complex problem with many partial solutions},''
  \emph{\BIBforeignlanguage{en}{J. Magn. Reson. Imaging}}, vol.~42, no.~4, pp.
  887--901, 2015.

\bibitem{Nguyen2022-yc}
X.~V. Nguyen \emph{et~al.}, ``Cost economy of motion,'' in \emph{Advances in
  Magnetic Resonance Technology and Applications}.\hskip 1em plus 0.5em minus
  0.4em\relax Elsevier, 2022, pp. 25--34.

\bibitem{Seiberlich2020-xs}
N.~Seiberlich \emph{et~al.}, Eds., \emph{Quantitative magnetic resonance
  imaging: Volume 1}, ser. Advances in Magnetic Resonance Technology and
  Applications.\hskip 1em plus 0.5em minus 0.4em\relax San Diego, CA: Academic
  Press, 2020.

\bibitem{Feng2016-nu}
L.~Feng \emph{et~al.}, ``\BIBforeignlanguage{en}{{XD}-{GRASP}: Golden-angle
  radial {MRI} with reconstruction of extra motion-state dimensions using
  compressed sensing},'' \emph{\BIBforeignlanguage{en}{Magn. Reson. Med.}},
  vol.~75, no.~2, pp. 775--788, 2016.

\bibitem{Towfighi2020-wu}
A.~Towfighi \emph{et~al.}, ``\BIBforeignlanguage{en}{Bridging the gap between
  research, policy, and practice: Lessons learned from academic-public
  partnerships in the {CTSA} network},'' \emph{\BIBforeignlanguage{en}{J. Clin.
  Transl. Sci.}}, vol.~4, no.~3, pp. 201--208, 2020.

\bibitem{Unknown2024-eh}
``\BIBforeignlanguage{en}{What will it take to make precision health a global
  reality},'' \emph{\BIBforeignlanguage{en}{Nat. Med.}}, vol.~30, no.~7, pp.
  1793--1794, 2024.

\bibitem{Rao2025-df}
V.~M. Rao \emph{et~al.}, ``\BIBforeignlanguage{en}{Multimodal generative {AI}
  for medical image interpretation},'' \emph{\BIBforeignlanguage{en}{Nature}},
  vol. 639, no. 8056, pp. 888--896, 2025.

\bibitem{Hagiwara2023-kr}
A.~Hagiwara \emph{et~al.}, ``\BIBforeignlanguage{en}{Multiparametric {MRI}:
  From simultaneous rapid acquisition methods and analysis techniques using
  scoring, machine learning, radiomics, and deep learning to the generation of
  novel metrics},'' \emph{\BIBforeignlanguage{en}{Invest. Radiol.}}, 2023.

\bibitem{van-Timmeren2020-ar}
J.~E. van Timmeren \emph{et~al.}, ``\BIBforeignlanguage{en}{Radiomics in
  medical imaging-``how-to'' guide and critical reflection},''
  \emph{\BIBforeignlanguage{en}{Insights Imaging}}, vol.~11, no.~1, p.~91,
  2020.

\bibitem{Hagiwara2020-lr}
A.~Hagiwara \emph{et~al.}, ``\BIBforeignlanguage{en}{Variability and
  standardization of quantitative imaging: Monoparametric to multiparametric
  quantification, radiomics, and artificial intelligence},''
  \emph{\BIBforeignlanguage{en}{Invest. Radiol.}}, vol.~55, no.~9, pp.
  601--616, 2020.

\bibitem{Paverd2024-tm}
H.~Paverd \emph{et~al.}, ``\BIBforeignlanguage{en}{Radiology and multi-scale
  data integration for precision oncology},'' \emph{\BIBforeignlanguage{en}{NPJ
  Precis. Oncol.}}, vol.~8, no.~1, p. 158, Jul. 2024.

\bibitem{yvonnelui_review}
D.~J. Lin \emph{et~al.}, ``Artificial intelligence for {MR} image
  reconstruction: An overview for clinicians,'' \emph{J. Magn. Reson. Imaging},
  vol.~53, no.~4, pp. 1015--1028, 2021.

\bibitem{Conti2024-am}
I.~Conti \emph{et~al.}, ``\BIBforeignlanguage{en}{Magnetic resonance imaging in
  pediatric epilepsy patients: two different protocols},''
  \emph{\BIBforeignlanguage{en}{J. Med. Imaging Interv. Radiol.}}, vol.~11,
  no.~1, 2024.

\bibitem{Flores2023-bi}
D.~V. Flores \emph{et~al.}, ``\BIBforeignlanguage{en}{Distal radioulnar joint:
  Normal anatomy, imaging of common disorders, and injury classification},''
  \emph{\BIBforeignlanguage{en}{Radiographics}}, vol.~43, no.~1, p. e220109,
  Jan. 2023.

\bibitem{johnson2023deep}
P.~M. Johnson \emph{et~al.}, ``Deep learning reconstruction enables
  prospectively accelerated clinical knee {MRI},'' \emph{Radiology}, vol. 307,
  no.~2, p. e220425, 2023.

\bibitem{Purysko2020-mo}
A.~S. Purysko \emph{et~al.}, ``\BIBforeignlanguage{en}{{RadioGraphics} update:
  {PI}-{RADS} version 2.1-a pictorial update},''
  \emph{\BIBforeignlanguage{en}{Radiographics}}, vol.~40, no.~7, pp. E33--E37,
  2020.

\bibitem{Rajiah2023-mz}
P.~S. Rajiah, C.~J. François, and T.~Leiner, ``\BIBforeignlanguage{en}{Cardiac
  {MRI}: State of the art},'' \emph{\BIBforeignlanguage{en}{Radiology}}, vol.
  307, no.~3, p. e223008, 2023.

\bibitem{oscanoa2023deep}
J.~A. Oscanoa \emph{et~al.}, ``Deep learning-based reconstruction for cardiac
  {MRI}: a review,'' \emph{Bioengineering}, vol.~10, no.~3, p. 334, 2023.

\bibitem{morales2024present}
M.~A. Morales, W.~J. Manning, and R.~Nezafat, ``Present and future innovations
  in {AI} and cardiac {MRI},'' \emph{Radiology}, vol. 310, no.~1, p. e231269,
  2024.

\bibitem{lecouvet2024present}
F.~E. Lecouvet \emph{et~al.}, ``{Present and future of whole-body MRI in
  metastatic disease and myeloma: How and why you will do it},'' \emph{Skeletal
  Radiol.}, vol.~53, no.~9, pp. 1815--1831, 2024.

\bibitem{Pollard2017-wj}
J.~M. Pollard \emph{et~al.}, ``\BIBforeignlanguage{en}{The future of
  image-guided radiotherapy will be {MR} guided},''
  \emph{\BIBforeignlanguage{en}{Br. J. Radiol.}}, vol.~90, no. 1073, p.
  20160667, 2017.

\bibitem{De-Pietro2024-mk}
S.~De~Pietro \emph{et~al.}, ``\BIBforeignlanguage{en}{The role of {MRI} in
  radiotherapy planning: a narrative review ``from head to toe''},''
  \emph{\BIBforeignlanguage{en}{Insights Imaging}}, vol.~15, no.~1, p. 255,
  2024.

\bibitem{Otazo2021-no}
R.~Otazo \emph{et~al.}, ``\BIBforeignlanguage{en}{{MRI}-guided radiation
  therapy: An emerging paradigm in adaptive radiation oncology},''
  \emph{\BIBforeignlanguage{en}{Radiology}}, vol. 298, no.~2, pp. 248--260,
  2021.

\bibitem{Thompson2021-sr}
S.~M. Thompson \emph{et~al.}, ``\BIBforeignlanguage{en}{Body interventional
  {MRI} for diagnostic and interventional radiologists: Current practice and
  future prospects},'' \emph{\BIBforeignlanguage{en}{Radiographics}}, vol.~41,
  no.~6, pp. 1785--1801, 2021.

\bibitem{Campbell-Washburn2019-ub}
A.~E. Campbell-Washburn \emph{et~al.}, ``\BIBforeignlanguage{en}{Opportunities
  in interventional and diagnostic imaging by using high-performance
  low-field-strength {MRI}},'' \emph{\BIBforeignlanguage{en}{Radiology}}, vol.
  293, no.~2, pp. 384--393, 2019.

\bibitem{shimronlowfield}
E.~Shimron \emph{et~al.}, ``{Accelerating Low-field MRI: Compressed Sensing and
  AI for fast noise-robust imaging},'' \emph{arXiv:2411.06704}, 2024.

\bibitem{Gannon2014-az}
F.~Gannon, ``\BIBforeignlanguage{en}{The steps from translatable to
  translational research},'' \emph{\BIBforeignlanguage{en}{EMBO Rep.}},
  vol.~15, no.~11, pp. 1107--1108, 2014.

\bibitem{fort2017mapping}
D.~G. Fort \emph{et~al.}, ``Mapping the evolving definitions of translational
  research,'' \emph{J. Clin. Transl. Sci.}, vol.~1, no.~1, pp. 60--66, 2017.

\bibitem{Reisenhofer2018-ni}
R.~Reisenhofer \emph{et~al.}, ``\BIBforeignlanguage{en}{A {Haar} wavelet-based
  perceptual similarity index for image quality assessment},''
  \emph{\BIBforeignlanguage{en}{Signal Process. Image Commun.}}, vol.~61, pp.
  33--43, 2018.

\bibitem{Karner2025-gf}
C.~Karner \emph{et~al.}, ``Parameter choices in {HaarPSI} for {IQA} with
  medical images,'' in \emph{IEEE ISBI}.\hskip 1em plus 0.5em minus 0.4em\relax
  IEEE, 2025, pp. 1--5.

\bibitem{Knoll2020-de}
F.~Knoll \emph{et~al.}, ``\BIBforeignlanguage{en}{Advancing machine learning
  for {MR} image reconstruction with an open competition: Overview of the 2019
  {fastMRI} challenge},'' \emph{\BIBforeignlanguage{en}{Magn. Reson. Med.}},
  vol.~84, no.~6, pp. 3054--3070, 2020.

\bibitem{Center-for-Devices2024}
{Center for Devices and Radiological Health}, ``Methods and tools for effective
  postmarket monitoring of artificial intelligence ({AI})-enabled medical
  devices,''
  \href{https://www.fda.gov/medical-devices/medical-device-regulatory-science-research-programs-conducted-osel/methods-and-tools-effective-postmarket-monitoring-artificial-intelligence-ai-enabled-medical-devices}{\nolinkurl{https://www.fda.gov/.../ai-enabled-medical-devices}},
  2024, [Online]. Accessed on Apr. 29, 2025.

\bibitem{hallucination}
V.~Antun \emph{et~al.}, ``{On instabilities of deep learning in image
  reconstruction and the potential costs of AI},'' \emph{Proc. Natl. Acad.
  Sci.}, vol. 117, no.~48, pp. 30\,088--30\,095, 2020.

\bibitem{shimron2022implicit}
E.~Shimron \emph{et~al.}, ``Implicit data crimes: Machine learning bias arising
  from misuse of public data,'' \emph{Proc. Natl. Acad. Sci}, vol. 119, no.~13,
  p. e2117203119, 2022.

\bibitem{sriram2020grappanet}
A.~Sriram \emph{et~al.}, ``{GrappaNet: Combining parallel imaging with deep
  learning for multi-coil MRI reconstruction},'' in \emph{IEEE CVPR}, 2020, pp.
  14\,315--14\,322.

\bibitem{ISTA-NET}
J.~{Zhang} and B.~{Ghanem}, ``{ISTA-Net: I}nterpretable optimization-inspired
  deep network for image compressive sensing,'' in \emph{{IEEE CVPR}}, 2018,
  pp. 1828--1837.

\bibitem{Primal_dual}
J.~Cheng \emph{et~al.}, ``Model learning: {Primal} dual networks for fast {MR}
  imaging,'' in \emph{MICCAI}, 2019, pp. 21--29.

\bibitem{dar2021fewshot}
S.~U.~H. Dar \emph{et~al.}, ``{Parallel-stream fusion of scan-specific and
  scan-general priors for learning deep MRI reconstruction in low-data
  regimes},'' \emph{Comput. Biol. Med.}, vol. 167, p. 107610, 2023.

\bibitem{Aggarwal2020-nd}
H.~K. Aggarwal, M.~P. Mani, and M.~Jacob, ``{MoDL-MUSSELS: model-based deep
  learning for multishot sensitivity-encoded diffusion MRI},'' \emph{IEEE
  Trans. Med. Imaging}, vol.~39, no.~4, pp. 1268--1277, 2020.

\bibitem{WELLER2018118}
D.~S. Weller, M.~Salerno, and C.~H. Meyer, ``Content-aware compressive magnetic
  resonance image reconstruction,'' \emph{Magn. Reson. Imaging}, vol.~52, pp.
  118--130, 2018.

\bibitem{edupuganti2020uncertainty}
V.~Edupuganti \emph{et~al.}, ``Uncertainty quantification in deep {MRI}
  reconstruction,'' \emph{IEEE Trans. Med. Imaging}, vol.~40, no.~1, pp.
  239--250, 2020.

\bibitem{anastasiotmi2021}
S.~Bhadra \emph{et~al.}, ``On hallucinations in tomographic image
  reconstruction,'' \emph{IEEE Trans. Med. Imaging}, vol.~40, no.~11, pp.
  3249--3260, 2021.

\bibitem{hallumetric}
C.-C. Chan and J.~P. Haldar, ``{Local perturbation responses and checkerboard
  tests: Characterization tools for nonlinear MRI methods},'' \emph{Magn.
  Reson. Med.}, vol.~86, no.~4, pp. 1873--1887, 2021.

\bibitem{bach2015pixel}
S.~Bach \emph{et~al.}, ``On pixel-wise explanations for non-linear classifier
  decisions by layer-wise relevance propagation,'' \emph{PloS one}, vol.~10,
  no.~7, p. e0130140, 2015.

\bibitem{bedel2023dreamr}
H.~A. Bedel and T.~Çukur, ``{DreaMR: Diffusion-driven} counterfactual
  explanation for functional {MRI},'' \emph{IEEE Trans. Med. Imaging}, 2024.

\bibitem{sundararajan2017axiomatic}
M.~Sundararajan, A.~Taly, and Q.~Yan, ``Axiomatic attribution for deep
  networks,'' in \emph{ICML}.\hskip 1em plus 0.5em minus 0.4em\relax PMLR,
  2017, pp. 3319--3328.

\bibitem{AIbias}
E.~A. Stanley, M.~Wilms, and N.~D. Forkert, ``{Disproportionate subgroup
  impacts and other challenges of fairness in artificial intelligence for
  medical image analysis},'' in \emph{EPIMI}.\hskip 1em plus 0.5em minus
  0.4em\relax Springer, 2022, pp. 14--25.

\bibitem{du2023}
Y.~Du \emph{et~al.}, ``Unveiling fairness biases in deep learning-based brain
  {MRI} reconstruction,'' in \emph{CLIP EPIMI FAIMI}, 2023, pp. 102--111.

\bibitem{guo2021}
P.~Guo \emph{et~al.}, ``{Multi-institutional collaborations for improving deep
  learning-based magnetic resonance image reconstruction using federated
  learning},'' in \emph{CVPR}, vol.~13, 2021, p. 2423–2432.

\bibitem{DINSDALE2021}
N.~K. Dinsdale, M.~Jenkinson, and A.~I. Namburete, ``Deep learning-based
  unlearning of dataset bias for {MRI} harmonisation and confound removal,''
  \emph{NeuroImage}, vol. 228, p. 117689, 2021.

\bibitem{dalmaz2024one}
O.~Dalmaz \emph{et~al.}, ``One model to unite them all: Personalized federated
  learning of multi-contrast {MRI} synthesis,'' \emph{Med. Image Anal.}, p.
  103121, 2024.

\bibitem{xu_addressing_2024}
Z.~Xu \emph{et~al.}, ``Addressing fairness issues in deep learning-based
  medical image analysis: a systematic review,'' \emph{NPJ Digit. Med.},
  vol.~7, no.~1, p. 286, 2024.

\bibitem{liu2019santis}
F.~Liu \emph{et~al.}, ``{SANTIS: sampling-augmented neural network with
  incoherent structure for MR image reconstruction},'' \emph{Magn. Reson.
  Med.}, vol.~82, no.~5, pp. 1890--1904, 2019.

\bibitem{han2018deep}
Y.~Han \emph{et~al.}, ``{Deep learning with domain adaptation for accelerated
  projection-reconstruction MR},'' \emph{Magn. Reson. Med.}, vol.~80, no.~3,
  pp. 1189--1205, 2018.

\bibitem{fastmri}
F.~Knoll \emph{et~al.}, ``{fastMRI: A} publicly available raw k-space and
  {DICOM} dataset of knee images for accelerated {MR} image reconstruction
  using machine learning,'' \emph{Radiol. Artif. Intell}, vol.~2, no.~1, p.
  e190007, 2020.

\bibitem{oh2020unpaired}
G.~Oh \emph{et~al.}, ``{Unpaired deep learning for accelerated MRI using
  optimal transport driven CycleGAN},'' \emph{IEEE Trans. Comput. Imaging},
  vol.~6, pp. 1285--1296, 2021.

\bibitem{lei2020wasserstein}
K.~Lei \emph{et~al.}, ``{Wasserstein GANs for MR imaging: from paired to
  unpaired training},'' \emph{IEEE Trans. Med. Imaging}, vol.~40, no.~1, pp.
  105--115, 2021.

\bibitem{ozbey2022unsupervised}
M.~Ozbey \emph{et~al.}, ``Unsupervised medical image translation with
  adversarial diffusion models,'' \emph{IEEE Trans. Med. Imaging}, vol.~42,
  no.~12, pp. 3524--3539, 2023.

\bibitem{elmas2022federated}
G.~Elmas \emph{et~al.}, ``Federated learning of generative image priors for
  {MRI} reconstruction,'' \emph{IEEE Trans. Med. Imaging}, vol.~42, no.~7, pp.
  1996--2009, 2023.

\bibitem{levac2023federated}
B.~R. Levac, M.~Arvinte, and J.~I. Tamir, ``{Federated end-to-end unrolled
  models for magnetic resonance image reconstruction},'' \emph{Bioengineering},
  vol.~10, no.~3, p. 364, 2023.

\bibitem{syntheticdata}
Z.~Wang \emph{et~al.}, ``{One for multiple: Physics-informed synthetic data
  boosts generalizable deep learning for fast MRI reconstruction},'' \emph{Med.
  Image Anal.}, p. 103616, 2025.

\bibitem{sun2025foundation}
Y.~Sun \emph{et~al.}, ``A foundation model for enhancing magnetic resonance
  images and downstream segmentation, registration and diagnostic tasks,''
  \emph{Nat. Biomed. Eng.}, vol.~9, no.~4, pp. 521--538, 2025.

\bibitem{dong2025foundation}
H.~Dong \emph{et~al.}, ``{MRI-CORE}: A foundation model for magnetic resonance
  imaging,'' \emph{arXiv:2506.12186}, 2025.

\bibitem{zhang2013coil}
T.~Zhang \emph{et~al.}, ``{Coil compression for accelerated imaging with
  Cartesian sampling},'' \emph{Magn. Reson. Med.}, vol.~69, no.~2, pp.
  571--582, 2013.

\bibitem{schaetz2017accelerated}
S.~Schaetz \emph{et~al.}, ``Accelerated computing in magnetic resonance
  imaging: real-time imaging using nonlinear inverse reconstruction,''
  \emph{Comput. Math. Methods Med.}, vol. 2017, no.~1, p. 3527269, 2017.

\end{thebibliography}

\end{document}